\documentclass[
reprint,
amsmath,
amssymb,
floatfix,
superscriptaddress,
twocolumn,
aps
]{revtex4-2}

\usepackage{amsmath,amssymb}
\usepackage{graphicx}
\usepackage{dcolumn}
\usepackage{bm}
\usepackage[dvipsnames]{xcolor} 
\usepackage{xfrac}
\usepackage{mathrsfs}
\usepackage{comment}
\usepackage{hyperref}
\hypersetup{
  colorlinks=true,
  linkcolor=Blue,
  citecolor=Blue,
  urlcolor=Blue
}

\newcommand{\bea}{\begin{eqnarray}}
\newcommand{\eea}{\end{eqnarray}}
\newcommand{\f}[2]{\frac{#1}{#2}}

\newcommand{\Gam}{\Gamma}
\newcommand{\la}{\langle}
\newcommand{\ra}{\rangle}
\newcommand{\nn}{\nonumber}
\newcommand{\p}{\partial}

\newcommand{\uv}{{\hat{\mathbf{u}}}}
\newcommand{\rv}{{\bf r}}
\newcommand{\fv}{{\bf F}}
\newcommand{\dr}{D_{r}}
\newcommand{\mupara}{\mu_{\parallel}}
\newcommand{\muperp}{\mu_{\perp}}
\newcommand{\beq}{\begin{equation}}
\newcommand{\eeq}{\end{equation}}

\makeatletter
\AtBeginDocument{\immediate\write\@mainaux{\string\@writefile{toc}{\string\iffalse}}}
\makeatother

\begin{document}

\title{Mobility Anisotropy Reshapes Self-Propelled Motion}

\author{Amir Shee}
\email[]{amir.shee@uvm.edu}
\affiliation{Department of Physics, University of Vermont,
Burlington, Vermont 05405, USA}

\author{P. S. Pal}
\email[]{pspal@kias.re.kr}
\affiliation{School of Computational Sciences, Korea Institute for
Advanced Study, Seoul 02455, Korea}

\begin{abstract}
We obtain exact closed-form moments for a harmonically trapped self-propelled particle with anisotropic translational mobility in two dimensions.
At high persistence, the mean and MSD develop a quasi-steady plateau with vanishing fluctuations, after which the MSD relaxes in a second step.
The steady excess kurtosis is negative and nonmonotonic in the mechanical-to-rotational timescale ratio, and the particle is pushed beyond the activity-confinement ring.
The non-monotonic kurtosis is specific to strong mobility anisotropy—as in wheeled or gliding particles, not bulk hydrodynamic rods—whereas the two-step MSD remains visible whenever transverse relaxation is slow compared with rotational diffusion.
\end{abstract}

\maketitle

Active and living matter comprises self-propelled units that convert microscale energy input into persistent directed motion~\cite{Ramaswamy2010}.
Examples range from motile microorganisms to synthetic systems such as catalytic nanomotors, light-activated swimmers, and engineered microrobots~\cite{Berg2004, Paxton2004, Tailleur2008, Jiang2010, Palacci2013, Goldstein2015, Dauchot2019}.
Continuous energy dissipation drives these systems intrinsically far from thermodynamic equilibrium, leading to emergent phenomena such as motility-induced phase separation, collective swarming, and anomalous transport~\cite{Marchetti2013}.
A canonical minimal framework is furnished by self-propelled particle (SPP) models~\cite{Romanczuk2012, Bechinger2016, Basu2018, Basu2019, Shee2020}, in which particles move at constant speed while their orientation diffuses rotationally.
Despite their simplicity, these models capture a broad range of experimental and simulation results in the bulk, near interfaces, and under confinement~\cite{Redner2013, Solon2015, Henkes2020}.

Many emergent phenomena in active matter can be captured by models of SPPs with \emph{isotropic} translational mobility.
In practice, however, physical constraints such as body shape, substrate interactions, or external fields often break this mobility symmetry.
Wheeled robots~\cite{Siegwart2011, Zheng2022}, bodies gliding on substrates, and particles in crowded or structured media can exhibit a pronounced mobility contrast, with transverse motion strongly suppressed relative to axial sliding~\cite{Elgeti2009, Elgeti2015, Bar2020, Peruani2006}.
By contrast, a slender rod in a bulk Newtonian fluid has only $\muperp/\mupara\simeq 1/2$~\cite{Doi1986, Peruani2006, Baskaran2008, Peruani2016}, too weak for the non-monotonic kurtosis derived below.
A systematic treatment of SPPs with anisotropic mobility is thus imperative.

The behavior of an isotropic SPP in harmonic confinement is well understood, with non-Boltzmann stationary distributions, circulating phase-space currents, and ring-like trajectories as characteristic signatures~\cite{Malakar2020, Chaudhuri2021}.
In contrast, the effect of anisotropic mobility on these nonequilibrium features remains unknown.
Strong anisotropy can confine the SPP to effectively one-dimensional motion along its preferred orientation, fundamentally altering the relaxation spectrum of the lower-order moments.
It moreover alters the higher-order statistics, thereby modifying the steady-state distribution.

We obtain exact closed-form moments for a minimal model of a self-propelled particle with anisotropic translational mobility in a harmonic trap.
We focus on the perfectly anisotropic limit, in which the mobility perpendicular to the self-propulsion direction is set to zero, and show in the End Matter that the rotationally invariant moment hierarchy closes for arbitrary transverse mobility, which quantifies when this limit applies.
Using a Laplace-transform method applied to the Fokker--Planck equation~\cite{Shee2020, Chaudhuri2021}, we derive closed-form time-dependent expressions for the mean displacement and the mean-squared displacement, together with their steady-state values.
Our results uncover a quasi-steady regime at intermediate times, emerging between the mechanical relaxation and rotational persistence timescales.
The anisotropic mobility tensor moreover introduces an irreducible quadratic structure into the moment hierarchy, with consequences that propagate to higher-order statistics.

We further derive exact steady-state fourth moments and characterize non-Gaussianity through the excess kurtosis.
Unlike the isotropic case~\cite{Malakar2020, Chaudhuri2021}, the anisotropic mobility yields a distinctive non-monotonic excess kurtosis, signaling qualitatively different steady-state position distributions.
The remainder of this Letter presents the model, the exact results, and their validation against numerical simulations.

\begin{figure}[!t]
\begin{center}
\includegraphics[width=\linewidth]{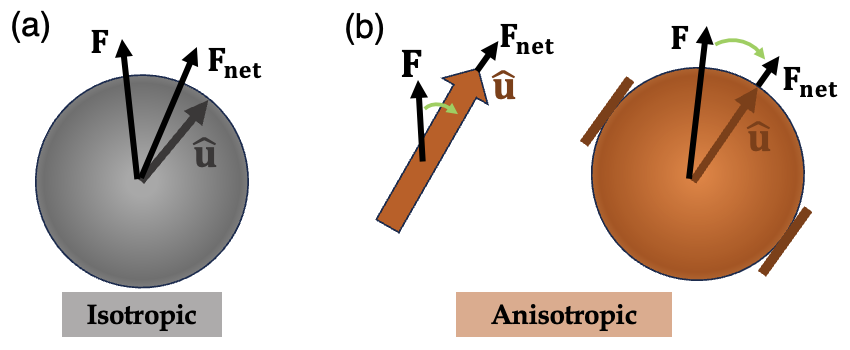}
\includegraphics[width=\linewidth]{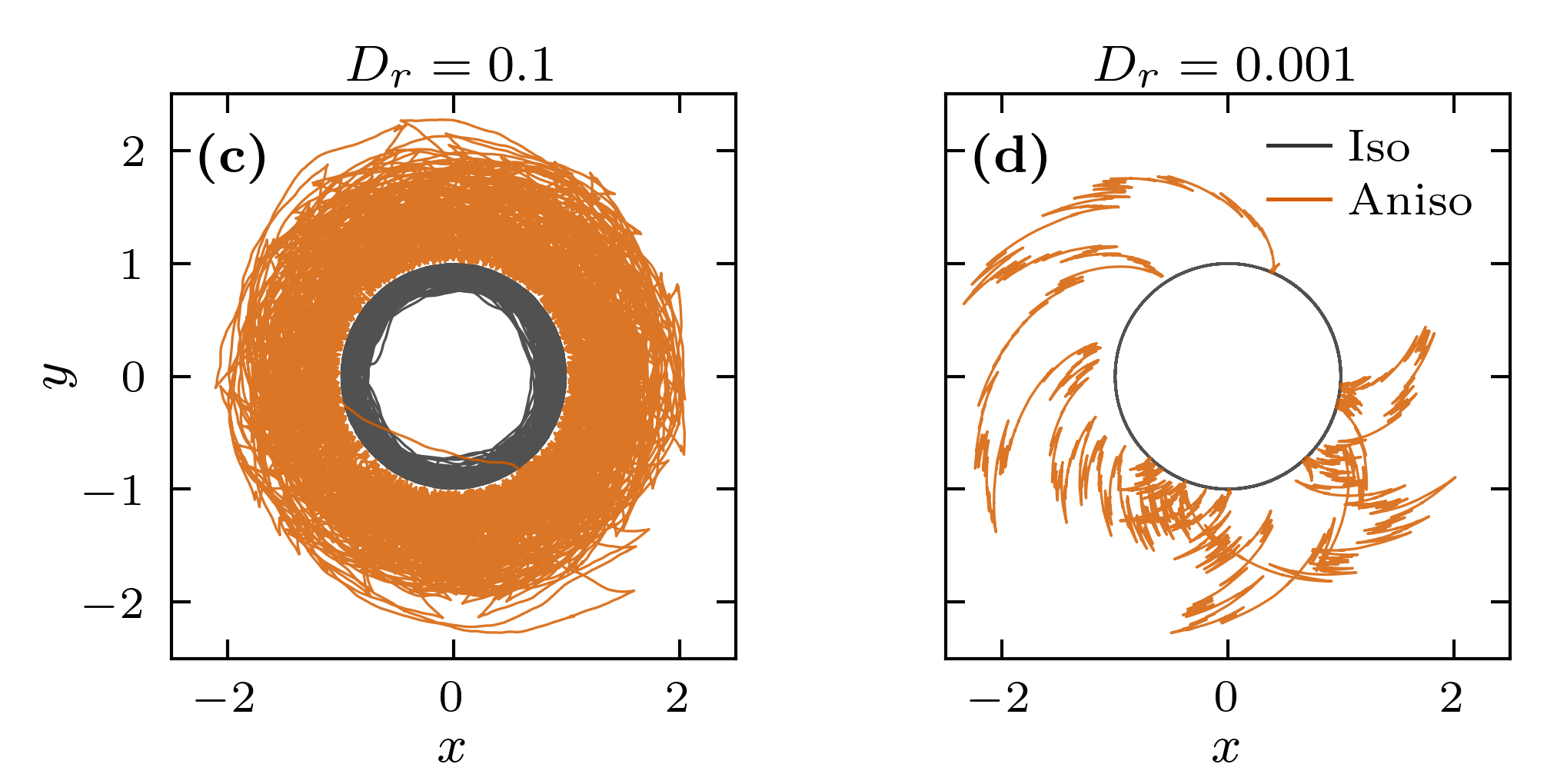}
\caption{
Self-propelled particle with isotropic and anisotropic mobility:
(a) For isotropic mobility ($\mu_\parallel=\mu_\perp=\mu$), the drift
induced by the external force $\mathbf{F}$ is collinear with the applied force and independent of the instantaneous orientation
$\hat{\mathbf{u}}$, implying that the net force $\mathbf{F}_{\rm net}$, defined as the vector sum of the self-propelled force and the external force, has a well-defined direction of its own.
(b) For anisotropic mobility ($\mu_\parallel\neq 0$, $\mu_\perp=0$),
only the force component along $\hat{\mathbf{u}}$ generates motion, so the net force is always directed along the instantaneous orientation $\uv$.
Representative steady-state trajectories for isotropic (gray)
and anisotropic (orange) mobility at (c) $D_r=0.1$ and (d)~$D_r=0.001$.
The self-propulsion speed $v_0=1$, trap stiffness $k=1$, and initial conditions: $\rv_0=\mathbf{0}$,
$\uv_0=\hat{\mathbf{x}}$.
Anisotropic mobility constrains the restoring force to the instantaneous propulsion direction, resulting in quasi-one-dimensional trajectories in the high-persistence limit.
}
\label{fig1}
\end{center}
\end{figure}

\medskip

\noindent
%
We consider an SPP at position $\rv$ that self-propels with constant speed $v_0$ along its orientation $\uv$.
The position dynamics with isotropic and anisotropic mobility reads
\bea
\dot{\rv} &=& v_0\,\uv + \mu\,\fv
  \qquad\qquad\quad\; ({\rm isotropic})\,,\label{eom1:disp}\\
\dot{\rv} &=& v_0\,\uv + \mu_{\parallel}\,\uv\uv^{T}\fv
  \qquad ({\rm anisotropic})\,,\label{eom2:disp}
\eea
where $\fv$ is the external force and $\uv\uv^T\fv=(\uv\cdot\fv)\uv$.
For isotropic mobility ($\mu$), the drift
is collinear with the applied force, with equal parallel and perpendicular components, $\mu_\parallel=\mu_\perp=\mu$ (Fig.~\ref{fig1}(a)).
In the strong anisotropic limit $\mu_\parallel\neq 0$, $\mu_\perp=0$,
the mobility tensor reduces to
$\boldsymbol{\mu}=\mu_\parallel\hat{\mathbf{u}}\hat{\mathbf{u}}^T$,
so only the force component along the body axis generates
motion (Fig.~\ref{fig1}(b)).
This limit is realized by non-holonomic (wheeled) robots, whose rolling constraint forbids lateral motion~\cite{Siegwart2011}, a platform widely used in collective-motion experiments~\cite{Zheng2022}, and is approached by elongated bodies whose transverse sliding is strongly suppressed by the substrate or the surrounding medium~\cite{Elgeti2015, Bar2020}.
For finite $\muperp$, the two-step relaxation of the MSD remains visible whenever transverse relaxation is slower than rotational relaxation, $\muperp k\lesssim D_r$; the narrower window in which the kurtosis approaches $-1/4$ is given in the End Matter.
A hydrodynamic rod, with $\muperp/\mupara\simeq 1/2$, lies outside this window whenever the orientation is persistent.
The orientation $\uv$ undergoes rotational diffusion,
\bea
\dot{\uv} = \sqrt{2D_r}\,\eta(t)\,\uv_\perp\,,
\label{eom3:rot}
\eea
where $\eta(t)$ is Gaussian white noise with
$\langle\eta(t)\eta(t')\rangle=\delta(t-t')$, giving the
standard orientational memory
$\langle\uv(t)\cdot\uv(0)\rangle=e^{-D_r t}$~\cite{Shee2020}.

The full statistical behavior is obtained from the probability distribution $P(\rv,\uv,t)$.
The Fokker--Planck equation for the
$P(\rv,\uv,t)$ with anisotropic mobility using Eq.~\eqref{eom2:disp} and Eq.~\eqref{eom3:rot} reads
\bea
\p_t P = \dr\nabla_{\uv}^2 P - v_0\,\uv\cdot\nabla P
       - \mupara\nabla\cdot(\uv\uv^T\fv P)\,.
\label{eq:F-P}
\eea
Applying the Laplace transform and integrating against an observable
$\psi(\rv,\uv)$ yields the moment-generator equation (see Supplemental Material, Sec.~\ref{SM:sec:aniso})
\bea
s\la\psi\ra_s &=& \la\psi\ra_0
  + \dr\la\nabla_{\uv}^2\psi\ra_s
  + v_0\la\uv\cdot\nabla\psi\ra_s\nn\\
&&+ \mupara\la\uv\uv^T\fv\cdot\nabla\psi\ra_s\,,
\label{eq:moment}
\eea
with $P(\rv,\uv,0)=\delta(\rv)\delta(\uv-\uv_0)$.
Eq.~\eqref{eq:moment} is the backbone for all moment
calculations below.
Note that the isotropic case follows by replacing $\mupara\uv\uv^T\fv \to \mu\fv$ in Eq.~\eqref{eq:F-P} and Eq.~\eqref{eq:moment}.

\begin{figure}[!t]
\begin{center}
\includegraphics[width=\linewidth]{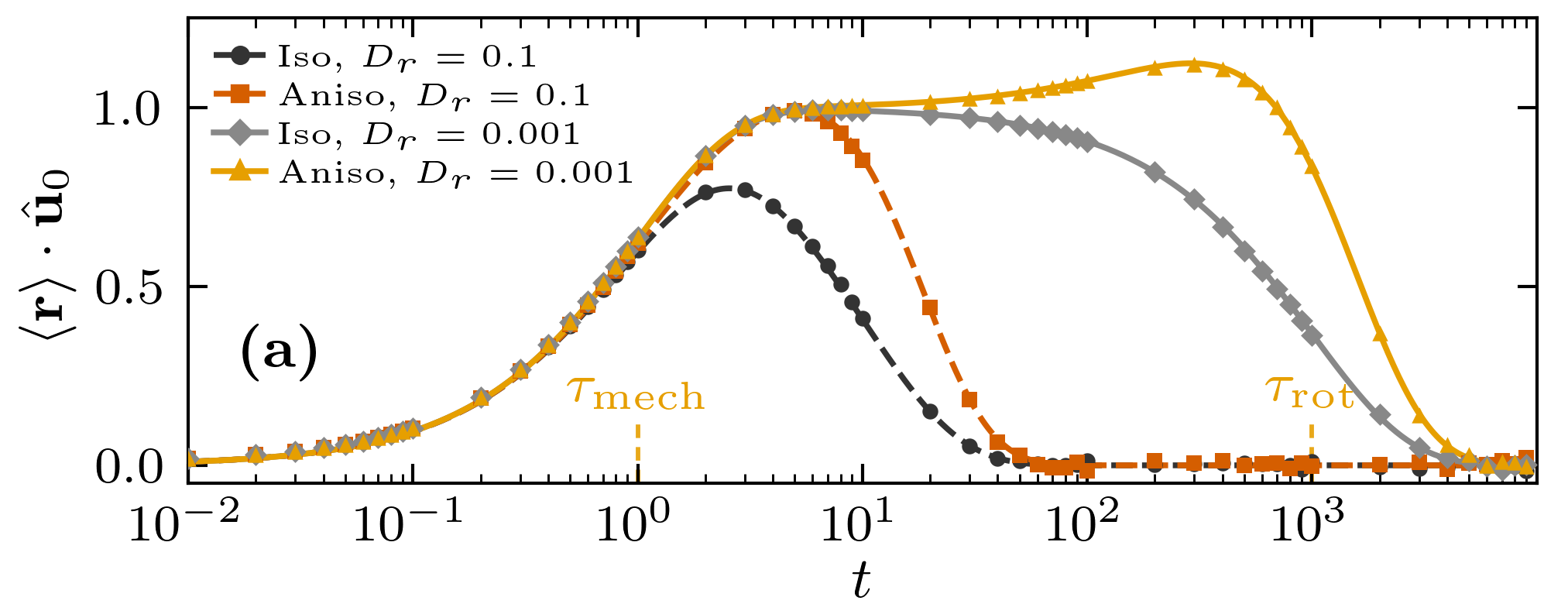}
\includegraphics[width=\linewidth]{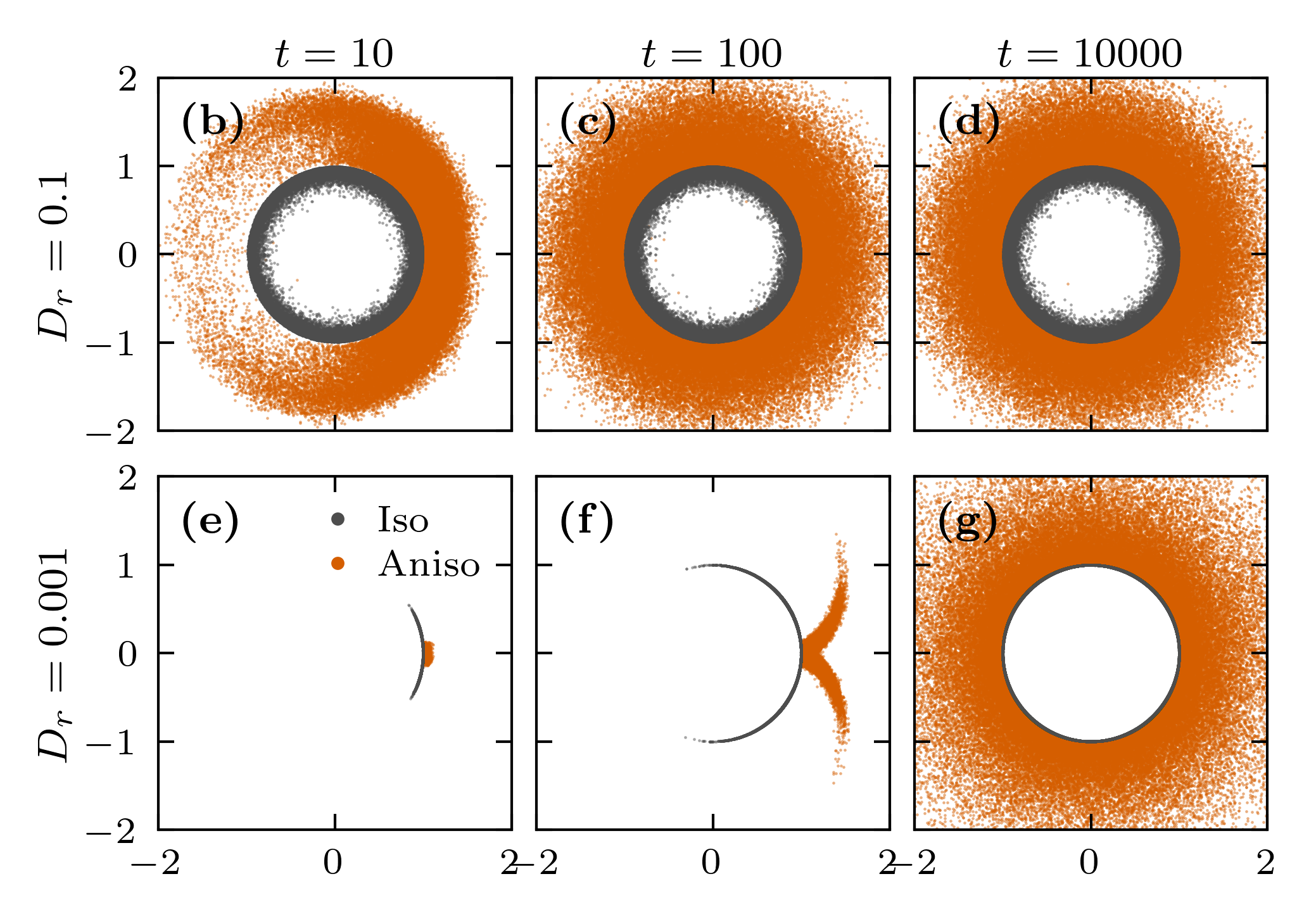}
\caption{
(a) Mean displacement along initial orientation $\langle\rv\rangle\cdot\uv_0$
for isotropic (Eq.~\eqref{eq:rav_isotropic}, End Matter) and anisotropic (Eq.~\eqref{eq:rav_anisotropic}) mobility at $D_r=0.1$ and $D_r=0.001$.
Curves are exact analytic results; symbols are simulation data.
Time evolution of the particle position distribution in a
harmonic trap for isotropic mobility ($\mu_\parallel=\mu_\perp=1$,
gray) and anisotropic mobility ($\mu_\parallel=1$, $\mu_\perp=0$,
orange) at $D_r=0.1$ (b-d) and $D_r=0.001$ (e-g).
Parameters: $v_0=1$, $k=1$, $\rv_{0}=\mathbf{0}$,
$\uv_{0}=\hat{\mathbf{x}}$.
}
\label{fig2}
\end{center}
\end{figure}

\begin{figure}[!t]
\begin{center}
\includegraphics[width=\linewidth]{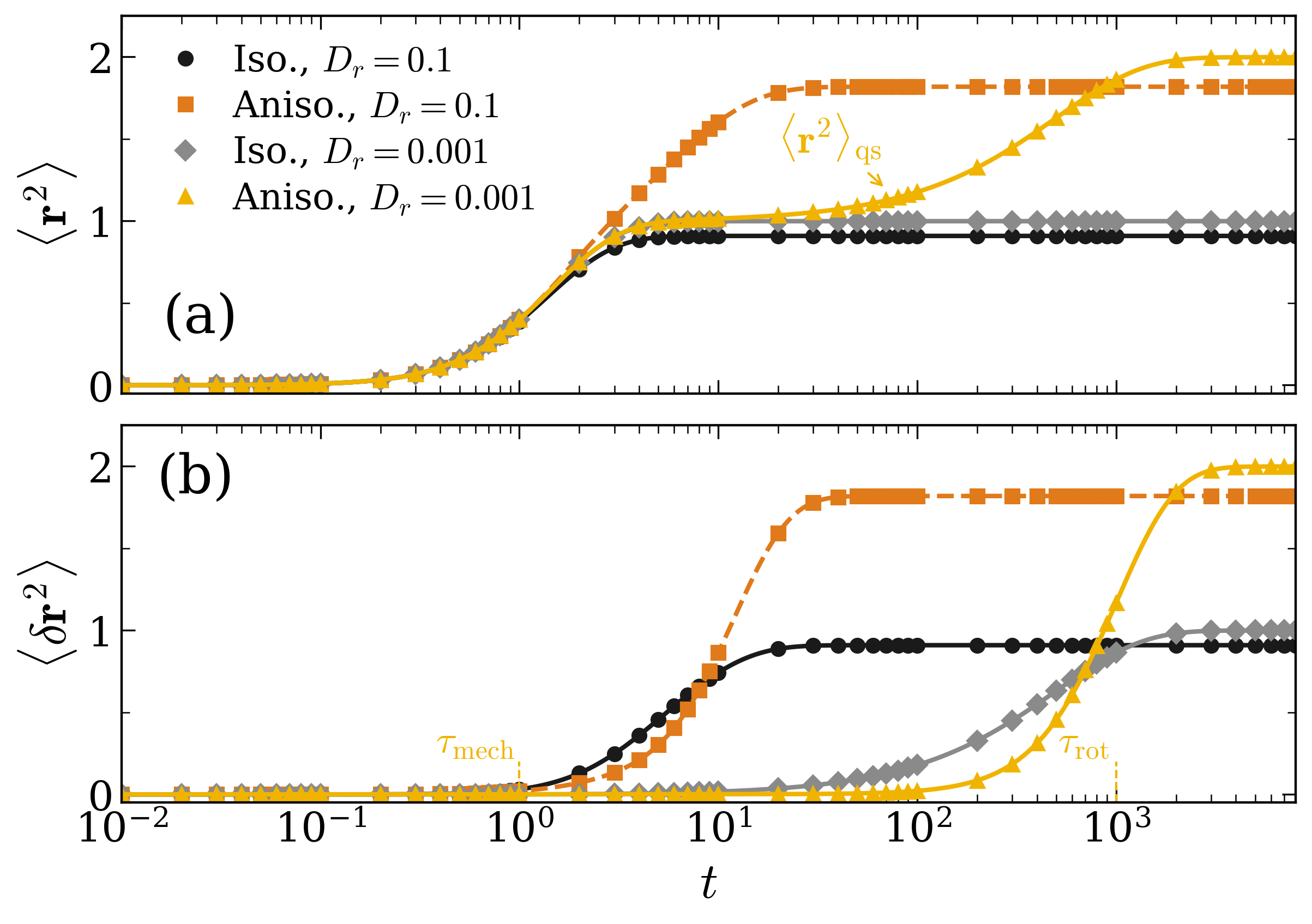}
\caption{Time evolution of the mean-squared displacement (MSD)
$\langle\rv^2\rangle$ (a) and displacement fluctuation $\la\delta \rv^2\ra$ (b) for isotropic ($\mu_\parallel=\mu_\perp=1$) and anisotropic ($\mu_\parallel=1$, $\mu_\perp=0$) mobility.
In the high-persistence limit ($D_r\ll\mu_\parallel k$), anisotropic
mobility produces a quasi-steady plateau
$\langle\rv^2\rangle_{\rm qs}=v_0^2/(\mu_\parallel k)^2$ between the mechanical timescale $\tau_{\rm mech}=1/\mu_\parallel k$
and the rotational timescale $\tau_{\rm rot}=1/D_r$, while the displacement fluctuation vanishes.
Curves are exact analytic results for isotropic (Eq.~\eqref{eq:r2av_isotropic} in End Matter) and anisotropic (Eq.~\eqref{eq:r2av_anisotropic}) mobility; symbols are simulation data.
Parameters: $v_0=1$, $k=1$, $\rv_0=\mathbf{0}$,
$\uv_0=\hat{\mathbf{x}}$.
}
\label{fig3}
\end{center}
\end{figure}

\noindent

Henceforth, we consider the external force to be harmonic, i.e., $\fv = - k \rv$.
The mean displacement of the SPP with anisotropic mobility (derivation in Supplemental Material, Sec.~\ref{SM:sec:aniso}) reads\footnote{Eq.~\eqref{eq:rav_anisotropic} is regular at $\mupara k=3\dr$, where the slow relaxation rate coincides with $\dr$. The limit is $\la\rv\ra=v_0\uv_0\big[\tfrac35\,t\,e^{-\dr t}+2\big(e^{-\dr t}-e^{-6\dr t}\big)/(25\dr)\big]$ (see Supplemental Material, Eq.~\eqref{SM:eq:rav_resonant}). The isotropic result, Eq.~\eqref{eq:rav_isotropic}, has the same removable singularity at $\mu k=\dr$.}
\begin{align}
\la\rv\ra &=
\frac{v_0\uv_0\,e^{-\left(4\dr+\mupara k
     -\sqrt{(4\dr)^2+(\mupara k)^2}\right)t/2}}
     {2(3\dr-\mupara k)\sqrt{(4\dr)^2+(\mupara k)^2}}
\nonumber\\
&\times\!\bigg\{
(12\dr+\mupara k)\!\left(1-e^{-\sqrt{(4\dr)^2+(\mupara k)^2}\,t}\right)
\nonumber\\
&+ 3\sqrt{(4\dr)^2+(\mupara k)^2}\,
\Big[1+e^{-\sqrt{(4\dr)^2+(\mupara k)^2}\,t}
\nonumber\\
&-2e^{\left(2\dr+\mupara k-\sqrt{(4\dr)^2+(\mupara k)^2}\right)t/2}
\Big]\bigg\}\,.
\label{eq:rav_anisotropic}
\end{align}
The square-root combination of $\dr$ and $\mupara k$ generates
multiple relaxation scales absent in the case of isotropic mobility~\cite{Chaudhuri2021} (see End Matter Eq.~\eqref{eq:rav_isotropic}).
In the frozen-orientation (i.e., high persistence) limit $\dr\to 0$,
$\la\rv\ra\to(v_0/\mupara k)(1-e^{-\mupara k t})\uv_0$, saturating to the quasi-steady value $\la\rv\ra_{\rm qs}=(v_0/\mupara k)\uv_0$ for $t\gtrsim\tau_{\rm mech}=1/\mupara k$.
The first signature of anisotropy in the small-time expansion appears at $\mathcal{O}(t^3)$ arising from the rotational Laplacian acting on $(\uv\cdot\rv)\uv$ (see Supplemental Material, Sec.~\ref{SM:sec:aniso}).

The mean parallel displacement $\langle \rv\rangle\cdot\uv_0$ is compared for both cases
in Fig.~\ref{fig2}(a).
At short times, all curves exhibit ballistic growth
$\langle\rv\rangle\cdot \uv_0 \simeq v_0 t$.
For isotropic mobility, the displacement relaxes smoothly to zero as rotational diffusion destroys the orientational memory.
In the high-persistence limit, both cases reach the quasi-steady value $a\equiv v_0/\mupara k$, where propulsion balances the trap along $\uv_0$.
The subsequent decay differs sharply.
Writing $\tau=\dr t$ and taking $\dr\ll\mupara k$, Eq.~\eqref{eq:rav_isotropic} gives $\la\rv\ra\cdot\uv_0=a\,e^{-\tau}$, whereas Eq.~\eqref{eq:rav_anisotropic} gives $\la\rv\ra\cdot\uv_0=a\,(3e^{-\tau}-2e^{-2\tau})$ [End Matter Eq.~\eqref{eq:slow_mean}].
The anisotropic mean therefore overshoots the plateau by exactly one eighth, reaching $9a/8$ at $\tau=\ln(4/3)$, before decaying to zero.
The contrasting spatial relaxation is illustrated in
Fig.~\ref{fig2}(b--g).
For isotropic mobility, the trap acts radially so the distribution relaxes to ring-like steady state rapidly.
For anisotropic mobility ($\muperp=0$), the trap acts only through
$(\uv\cdot\fv)\uv=-k(\uv\cdot\rv)\uv$.
Here, transverse displacements
are not directly restored and relax only indirectly via rotational
diffusion.
This produces strong elongated wing-shaped distributions at intermediate times where the
distribution collapses along $\hat{\mathbf{x}}$ before slow
rotational diffusion restores angular symmetry at
$t\sim\tau_{\rm rot}=1/D_r$ (see Fig.~\ref{fig2}).

\medskip

\noindent
The effect of timescale separation between $\tau_{\rm mech}$ and $\tau_{\rm rot}$ becomes more apparent when quantifying fluctuations via the mean-squared displacement (MSD).
The exact MSD for anisotropic mobility (derivation presented in Supplemental Material, Sec.~\ref{SM:sec:aniso}) reads
\begin{align}
\la\rv^{2}\ra
&= \frac{2v_0^2}{\mupara k(\dr+\mupara k)}
+ \frac{2v_0^2(3\dr-\mupara k)\,e^{-(\dr+\mupara k)t}}
       {(\dr+\mupara k)(3\dr^2+(\mupara k)^2)}\nn\\
&-\frac{2v_0^2\,e^{-(2\dr+\mupara k)t}}
       {\mupara k(3\dr^2+(\mupara k)^2)}
\bigg[3\dr\cosh\!\left(\sqrt{4\dr^2+(\mupara k)^2}\,t\right)
\nn\\
&\quad+\frac{6\dr^2+(\mupara k)^2}{\sqrt{4\dr^2+(\mupara k)^2}}
  \sinh\!\left(\sqrt{4\dr^2+(\mupara k)^2}\,t\right)\bigg]\,.
\label{eq:r2av_anisotropic}
\end{align}
The three transient contributions decay at rates
$(\dr+\mupara k)$ and
$(2\dr+\mupara k)\pm\sqrt{4\dr^2+(\mupara k)^2}$.
This square-root structure, absent in the isotropic result (see End Matter Eq.~\eqref{eq:r2av_isotropic}),
is the hallmark of anisotropic mobility.
Here, $\dr$ and $\mupara k$
enter the relaxation rates through a nonlinear coupling, in contrast to their independent superposition in the isotropic case.
In Fig.~\ref{fig3}(a), we compare Eq.~\eqref{eq:r2av_anisotropic} and End Matter Eq.~\eqref{eq:r2av_isotropic} against numerical simulations, finding excellent agreement.
The first signature of anisotropy in the small-time expansion of MSD appears at
$\mathcal{O}(t^4)$ (see Supplemental Material, Eq.~\eqref{SM:eq:r2av_anisotropic_small_time}).
The steady-state MSD is $\la\rv^2\ra_{\rm st}
= 2v_0^2/[\mupara k(\dr+\mupara k)]$, a factor of two larger than the isotropic value (see End Matter).
In the frozen-orientation limit $\dr\to 0$, the MSD saturates at the quasi-steady plateau $\la\rv^2\ra_{\rm qs}=v_0^2/(\mupara k)^2=a^2$ (see Supplemental Material, Sec.~\ref{SM:sec:aniso}).
This is identical to the isotropic steady-state value in the high-persistence limit (Fig.~\ref{fig3}(a)).
In the same limit, the anisotropic SPP approaches the quasi-steady mean position $a$ along $\uv_0$, which equals the isotropic steady-state mean position.
In this quasi-steady regime, $\tau_{\rm mech}< t < \tau_{\rm rot}$, the displacement fluctuation $\la\delta\rv^2\ra$ vanishes (Fig.~\ref{fig3}(b)).
On the other hand, for $t \gtrsim \tau_{\rm rot}$, $\la\delta\rv^2\ra$ increases and saturates to $\la\rv^2\ra_{\rm st}$.
It encodes a dimensional crossover at high persistence regime: from effectively one-dimensional, orientation-constrained motion at intermediate times to fully isotropic confined dynamics at long times.

The plateau has a simple origin.
For frozen orientation, the particle moves only along $\uv$ and stops where propulsion balances the projected trap force, $\uv\cdot\rv=a$, while its transverse coordinate is not restored.
The fixed points thus form a neutral line rather than an isolated point.
Rotational diffusion displaces the particle along this line, so the transverse coordinate performs an Ornstein--Uhlenbeck process with variance $a^2$ [End Matter Eq.~\eqref{eq:bodyframe}], and the MSD relaxes from $a^2$ to $2a^2$ at the single rate $2\dr$ [End Matter Eq.~\eqref{eq:slow_msd}].
In the isotropic case the same rotation moves the particle rigidly along an attracting ring, and the MSD does not change.
The two-step relaxation of the MSD is therefore the dynamical signature specific to anisotropy.

\medskip

\begin{figure*}[!t]
\begin{center}
\includegraphics[width=\linewidth]{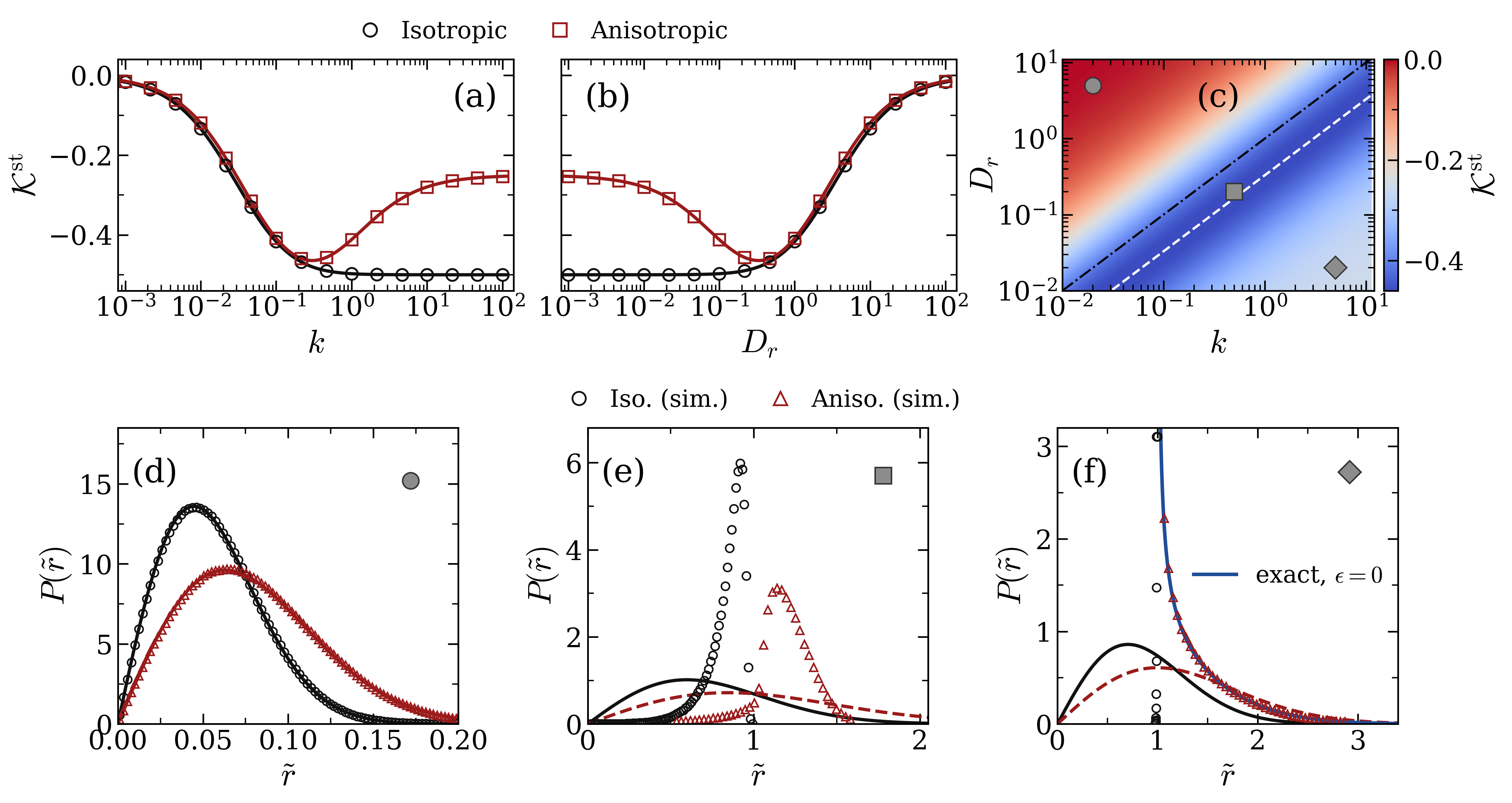}
\caption{
Non-Gaussian statistics:
Steady-state excess kurtosis $\mathcal{K}^{\rm st}$
versus trap stiffness $k$ at $D_r=0.1$ (a) and rotational diffusivity
$D_r$ at $k=1$ (b).
Solid lines are analytical predictions for isotropic (Eq.~\eqref{eq:alpha2_iso} in End Matter) and anisotropic (Eq.~\eqref{eq:alpha2_anisotropic}) mobility; open symbols are simulation
data.
(c)~Heat map of $\mathcal{K}^{\rm st}$ for the anisotropic case in the
$(k,D_r)$ plane.
Lines marked $\chi=1$ (black, dashed-dotted) and $\chi^*\simeq0.332$ (white, dashed).
(d)--(f)~Scaled radial distributions $P(\tilde r)$, $\tilde r=r\mu_\parallel k/v_0$, at the parameter points marked in
(c).
The lines are analytic Rayleigh radial distributions plotted using the steady-state MSD;   the solid blue curve in (f) is the exact high-persistence distribution, End Matter Eq.~\eqref{eq:P_hp} , at $\epsilon=\muperp k/D_r=0$.
}
\label{fig4}
\end{center}
\end{figure*}

\noindent
%
We characterize the shape of the position distribution through the excess kurtosis $\mathcal{K}=\la\rv^4\ra/(2\la\rv^2\ra^2)-1$.
The moment hierarchy closes at every order (End Matter), so $\la\rv^4\ra$ and the Cartesian components $\la x^2\ra$ and $\la y^2\ra$ are available in closed form in Laplace space (Supplemental Material, Sec.~\ref{SM:sec:aniso}). However, here we focus on the steady state.
The exact steady-state fourth moment (derivation presented in Supplemental Material, Sec.~\ref{SM:sec:aniso}) reads
\bea
\la\rv^4\ra_{\rm st}
&=& \frac{2v_0^4(48\dr^2+20\dr\mupara k+3(\mupara k)^2)}
         {(\mupara k)^2(4\dr+\mupara k)(\dr+\mupara k)}\nn\\
&&~~~~~~\times\frac{1}{3\dr^2+8\dr\mupara k+(\mupara k)^2}\,.
\label{eq:r4_aniso_st}
\eea
The corresponding non-Gaussian parameter, $\mathcal{K}^{\rm st}=\lim_{t\to\infty}\mathcal{K}$, is
\bea
\mathcal{K}^{\rm st}
&=& -\,\frac{\mupara k(72\dr^2+25\dr\mupara k+(\mupara k)^2)}
            {4(4\dr+\mupara k)(3\dr^2+8\dr\mupara k+(\mupara k)^2)}\,.
\label{eq:alpha2_anisotropic}
\eea
which is strictly non-positive, confirming sub-Gaussian steady states.
The steady-state fourth moment of displacement and excess kurtosis, for the isotropic case, are presented in End Matter Eqs.~\eqref{eq:r4_iso_st} and~\eqref{eq:alpha2_iso}, respectively.
A complete characterization of the steady-state non-Gaussian behavior is presented in Fig.~\ref{fig4}.
We plot $\mathcal{K}^{\rm st}$ as a function of $k$ (Fig.~\ref{fig4}(a)) and  $D_r$ (Fig.~\ref{fig4}(b)), showing excellent agreement between analytical predictions and simulation.
Unlike the isotropic case, the anisotropic $\mathcal{K}^{\rm st}$ exhibits non-monotonic behavior with respect to both $k$ and $D_r$.
This behavior is more transparently captured by the ratio $\chi=(D_r/\mupara k)(=\tau_{\rm mech}/\tau_{\rm rot})$ and it implies the existence of a critical ratio $\chi^*$ at which the non-Gaussianity is maximized.
Setting $d\mathcal{K}^{\rm st}/d\chi=0$ in Eq.~\eqref{eq:alpha2_anisotropic} gives
\bea
864\chi^4+600\chi^3+47\chi^2-74\chi-13=0\,,
\label{eq:chistar}
\eea
whose unique positive root is $\chi^*\simeq0.332$, with $\mathcal{K}^{\rm st,*}\simeq-0.464$.
Although $\chi^*$ lies close to $1/3$, where the prefactor of Eq.~\eqref{eq:rav_anisotropic} is singular, the two are unrelated: the quartic does not vanish at $\chi=1/3$, and the two points move apart for finite $\muperp$ (End Matter).
The limiting behaviors are as follows: in the high-persistence or strong-trapping limit, the steady-state excess kurtosis $\mathcal{K}^{\rm st}$ approaches $-1/2$ in the isotropic case and $-1/4$ in the anisotropic case (see Fig.~\ref{fig4}(a,b)).
In the opposite, low-persistence or weak-trapping limit, both isotropic and anisotropic $\mathcal{K}^{\rm st}$ vanish (Fig.~\ref{fig4}(a,b)).

In Fig.~\ref{fig4}(c), we plot a heat map of the non-Gaussianity on the $k-D_r$ plane.
Three regimes emerge: a near-Gaussian regime at $\chi\gg\chi^*$, and two active regimes, namely active-I near $\chi^*$ and active-II at $\chi\ll\chi^*$.
Three points, in the near-Gaussian regime, near $\chi^*$, and in the high-persistence regime, are marked on the $k-D_r$ plane to illustrate deviations of the simulated steady-state position distributions from the analytic Rayleigh radial distribution: $P(\tilde{r})=2\tilde{r}/\la\tilde{\rv}^2\ra_{\rm st} \exp(-\tilde{r}^2/\la\tilde{\rv}^2\ra_{\rm st})$ where $\tilde{r}=r/a$ and $\la\tilde{\rv}^2\ra_{\rm st}=\la\rv^2\ra_{\rm st}/a^2$ with $a=v_0/\mupara k $.
In Fig.~\ref{fig4}(d), the case $\mathcal{K}^{\rm st}\simeq 0$ yields a near-Gaussian steady-state position distribution, in close agreement with the analytic curves.

Figs.~\ref{fig4}(e) and \ref{fig4}(f) show the active-I and active-II regimes. Panel (e) is at $\chi=0.4$, where $\mathcal{K}^{\rm st}\simeq-0.463$, close to $\mathcal{K}^{\rm st,*}\simeq-0.464$. Panel (f) is at $\chi=0.004$, where $\mathcal{K}^{\rm st}\simeq-0.263$, approaching the limiting value $-1/4$.
These two active regimes are distinguished on the basis of the following features: (i) the tail of the distribution, (ii) the peak position of the distribution, and (iii) the distribution for $\tilde r < 1$.
In Fig.~\ref{fig4}(e), the steady-state position distribution exhibits a pronounced light-tailed profile in both cases, with the isotropic case being more prominent.
The most probable position (peak of the distribution) of the SPP in the anisotropic case lies outside the ring of radius $\tilde r=1$, whereas it lies inside for the isotropic case.
For isotropic mobility the entire position distribution is confined to $\tilde r\leq 1$ at all times and for any $\dr$, since $d|\rv|/dt=v_0(\uv\cdot\hat\rv)-\mu k|\rv|\leq v_0-\mu k|\rv|$.
No such bound exists for $\muperp=0$, because the transverse coordinate is not restored, and the dominant contribution to the anisotropic distribution arises from $\tilde r>1$.
In the high-persistence, strong-trapping limit, the anisotropy forces the particle into the high-potential region and $\mathcal{K}^{\rm st}\to-1/4$. Fig.~\ref{fig4}(f), at $\chi=0.004$, already lies close to this limit.
In this limit the longitudinal coordinate $\ell=\uv\cdot\rv$ is pinned at $\tilde\ell=\ell/a=1$, while the transverse coordinate $c=\uv_\perp\cdot\rv$  is Gaussian  of unit variance, so $\tilde r^2=1+\tilde c^2$ with $\tilde c=c/a$. This yields the exact distribution of End Matter Eq.~\eqref{eq:P_hp} without fitting parameters.
This is quite counterintuitive compared to the isotropic case, where the steady-state position distribution collapses to a delta-ring at radius $\tilde r = 1$ with $\mathcal{K}^{\rm st}=-1/2$.
In this parameter regime the region $\tilde r<1$ carries vanishing weight in both cases; for the anisotropic case the depletion is exact only as $\chi\to0$, and is smoothed at finite $\chi$ by the fluctuations of the longitudinal coordinate.

\medskip

\noindent
%
In conclusion, we have obtained exact closed-form moments for a harmonically trapped, self-propelled particle with anisotropic translational mobility, including time-dependent expressions and exact steady-state statistics.
In the high-persistence regime, the mean displacement and the MSD display a quasi-steady plateau with vanishing displacement fluctuations, after which the MSD relaxes in a second step to twice the plateau value.
The MSD exhibits a dimensional crossover from one-dimensional, orientation-constrained motion during the quasi-steady state to fully isotropic confined dynamics at long times.
The quasi-steady state emerges when the mechanical relaxation time is much shorter than the persistence time.

We characterize the steady state by exactly calculating the excess kurtosis for anisotropic mobility.
The steady state excess kurtosis shows non-monotonic behavior with both trap strength and rotational diffusion.
At high-persistence regime, it saturates to $-0.25$ for anisotropic case unlike $-0.50$  for isotropic.
Thus, the steady state position distribution shows fundamentally different non-Gaussian behavior compared to isotropic mobility.
Unlike the isotropic case, the steady-state position of an anisotropic particle can extend beyond the ring determined by the self-propulsion speed and the mechanical relaxation time.
For finite transverse mobility, the two-step MSD and the enhanced steady-state MSD remain pronounced whenever $\muperp k\lesssim\dr$, and the kurtosis minimum persists for $\lambda<\lambda_c\simeq0.266$, with $\mathcal{K}^{{\rm st},*}\simeq-0.464$ at $\muperp=0$; the minimum deepens toward $-1/2$ as $\lambda\to\lambda_c$ (Supplemental Material, Table~\ref{SM:tab:chistar}).
The saturation value $-1/4$ requires much stronger anisotropy: the closest approach to it deviates from $-1/4$ by an amount proportional to $\sqrt{\muperp/\mupara}$, so resolving it requires $\muperp/\mupara\lesssim10^{-4}$, whereas $\muperp k\lesssim\dr$ already keeps the high-persistence value above $-7/18\simeq-0.389$, well separated from the isotropic $-1/2$ (End Matter).

Our predictions apply when the confinement is harmonic over the ring radius $a=v_0/(\mupara k)$, the particle is small compared with $a$, and self-alignment is weak, that is, the orientation relaxes towards the velocity on a timescale longer than $\tau_{\rm rot}$; in walkers and robots where this is not the case, an additional self-aligning torque must be retained~\cite{Dauchot2019}.
The non-monotonic kurtosis requires $\lambda<\lambda_c\simeq0.266$, stronger than the hydrodynamic-rod value $\lambda\simeq 1/2$~\cite{Doi1986}; the two-step MSD requires $\muperp k\lesssim\dr$.
Wheeled robots in curved arenas supply both a soft gravitational trap and $\muperp\simeq0$, so that $\muperp k\lesssim D_r$ holds for any accessible persistence, provided self-alignment is weak compared with $\tau_{\rm rot}$~\cite{Siegwart2011, Zheng2022}; when self-alignment dominates, as in some wheeled walkers, an aligning torque must be retained~\cite{Dauchot2019}.
Elastically tethered active units provide a harmonic restoring force on the same footing, provided $a$ remains within the linear spring range~\cite{Baconnier2022}.
Acoustic or extended optical traps can furnish the required $a$~\cite{Takatori2016, Buttinoni2022}, but only if the particle itself satisfies $\lambda<\lambda_c\simeq0.266$; a micron-sized sphere cannot.

\medskip

\noindent
\textbf{Acknowledgments-}
A.S.\ thanks Cristi\'{a}n Huepe for valuable discussions on anisotropic mobility.
P.S.P.\ acknowledges research support from the Korea Institute for
Advanced Study through individual KIAS Grant No.~CG085601.

\medskip

\noindent
\textbf{Data availability-}
The data that support the findings of this article are available within the article and its Supplemental Material.

\medskip

\noindent
\textbf{Code availability-}
Simulation and analysis codes are available from the corresponding author upon reasonable request.

\bibliography{reference}

\section*{End Matter}

\noindent
\textbf{Exact analytic results for isotropic mobility:~}
We provide the exact analytic expressions for the time-dependent moments and the steady-state excess kurtosis here for completeness.
The explicit derivation is provided in Chaudhuri \& Dhar~\cite{Chaudhuri2021} and the Supplemental Material, Sec.~\ref{SM:sec:iso}.

\medskip

\noindent
The exact mean displacement:
\bea
\la\rv\ra
= -\,\frac{v_0\uv_0(e^{-\dr t}-e^{-\mu k t})}{\dr-\mu k}\,.
\label{eq:rav_isotropic}
\eea

\medskip

\noindent
The exact MSD:
\bea
\la\rv^{2}\ra
&=& \frac{v_0^2}{\mu k(\dr+\mu k)}
- \frac{2v_0^2\,e^{-\mu k t}}{\dr-\mu k}
  \left[\frac{e^{-\mu k t}}{2\mu k}
       -\frac{e^{-\dr t}}{\dr+\mu k}\right]\,,\nonumber\\
\label{eq:r2av_isotropic}
\eea
The transient relaxation towards steady-state consists of two rates: $2\mu k$ and $(\mu k+ D_r)$.
The steady-state value
$\la\rv^2\ra_{\rm st}=v_0^2/[\mu k(\dr+\mu k)]$.

\medskip

\noindent
The exact steady-state fourth moment:
\bea
\la\rv^4\ra_{\rm st}
&=& \frac{v_0^4(4\dr+3\mu k)}
         {(\mu k)^2(\dr+3\mu k)(\dr+\mu k)(2\dr+\mu k)}\,.
\label{eq:r4_iso_st}
\eea

\medskip

\noindent
The non-Gaussian parameter (i.e., excess kurtosis) at steady state:
\bea
\mathcal{K}^{\rm st}
&=& -\,\frac{\mu k(7\dr+3\mu k)}
            {2(2\dr+\mu k)(\dr+3\mu k)}\,.
\label{eq:alpha2_iso}
\eea

\medskip

\noindent
\noindent
\textbf{Position-orientation cross-correlation:~}
A quantity directly accessible in single-particle
tracking experiments is $\la\uv\cdot\rv\ra$, the
projection of position onto the instantaneous
propulsion axis. Setting $\psi=\uv\cdot\rv$ in
Eq.~\eqref{eq:moment} yields an exact closed form
that depends only on the parallel rate
$(\dr+\mupara k)$, independently of $\muperp$:
\bea
\la\uv\cdot\rv\ra
= \frac{v_0}{\dr+\mupara k}
  \left(1-e^{-(\dr+\mupara k)t}\right)\,.
\label{eq:urav_anisotropic}
\eea
This form is \emph{independent of $\muperp$} and
identical to the isotropic result under $\mu\to\mupara$
(derivation in Supplemental Material, Secs.~\ref{SM:sec:aniso} and~\ref{SM:sec:general}).
The transient decay rate $(D_r+\mupara k)$ is therefore a direct
and selective observable: measuring the cross-correlation decay in
a single-particle tracking experiment in an optical or magnetic trap
yields $\mupara$ once $\dr$ is obtained from the orientational autocorrelation and $k$ from calibration, without contamination from $\muperp$.

\medskip

\noindent
\textbf{Finite transverse mobility:~}
For the general tensor $\boldsymbol{\mu}=\mupara\uv\uv^T+\muperp(\mathbb{I}-\uv\uv^T)$, we write $\xi=\mupara k$, $\zeta=\muperp k$, $\lambda=\muperp/\mupara$, and $a=v_0/\xi$.
The body-frame coordinates $\ell=\uv\cdot\rv$ and $c=\uv_\perp\cdot\rv$, with $r^2=\ell^2+c^2$, obey the It\^o equations (Supplemental Material, Sec.~\ref{SM:sec:general})
\bea
d\ell &=& \big[v_0-(\dr+\xi)\,\ell\big]\,dt+\sqrt{2\dr}\,c\,dW\,,\nn\\
dc &=& -(\dr+\zeta)\,c\,dt-\sqrt{2\dr}\,\ell\,dW\,.
\label{eq:bodyframe}
\eea
The drift is linear and the noise bilinear, so moments of degree $p$ in $(\ell,c)$ couple only to moments of degree $\le p$.
Since the dynamics also conserves the total angular harmonic of position and orientation, all rotationally invariant moments therefore close order by order for any $\muperp$;  so do the Cartesian components (Supplemental Material, Sec.~\ref{SM:sec:aniso}).
The dynamics of $\ell$ in Eq.~\eqref{eq:bodyframe} leads to Eq.~\eqref{eq:urav_anisotropic} directly, upon averaging.
At second order one finds
\bea
\la\rv^2\ra_{\rm st}=\frac{v_0^2\,(2\dr+\zeta)}{(\dr+\xi)\big[\dr(\xi+\zeta)+\xi\zeta\big]}\,,
\label{eq:r2_general}
\eea
which reduces to the anisotropic and isotropic results for $\zeta=0$ and $\zeta=\xi$.
For frozen orientation, $\ell\to a$ and $c$ stays at zero for any $\muperp$, so the quasi-steady plateau is unaffected.

\medskip

\noindent
\textbf{High-persistence scaling forms:~}
For $\dr,\zeta\ll\xi$ the longitudinal coordinate relaxes at rate $\dr+\xi\simeq\xi$, much faster than $c$ and $\uv$, and can be eliminated adiabatically. The exact Laplace-space moments for arbitrary $\muperp$ then reduce to the slow modes, giving, with $\epsilon\equiv\zeta/\dr$ (Supplemental Material, Sec.~\ref{SM:sec:general}),
\bea
\la\rv\ra\cdot\uv_0&=&\frac{a}{1+\epsilon}\Big[(3+\epsilon)\,e^{-\dr t}-2\,e^{-(2\dr+\zeta)t}\Big],
\label{eq:slow_mean}\\
\la\rv^{2}\ra&=&\frac{a^2}{1+\epsilon}\Big[(2+\epsilon)-e^{-2(\dr+\zeta)t}\Big].
\label{eq:slow_msd}
\eea
For $\epsilon=0$ the mean overshoots its plateau, peaking at $9a/8$ at $\dr t=\ln(4/3)$, and the MSD rises from $a^2$ to $2a^2$ at rate $2\dr$; for $\epsilon\to\infty$ the isotropic forms $a\,e^{-\dr t}$ and $a^2$ are recovered.
In the same limit $c$ is Gaussian with variance $a^2/(1+\epsilon)$, so that
\bea
\la\rv^2\ra_{\rm st}\to a^2\,\frac{2+\epsilon}{1+\epsilon}\,,\qquad
\mathcal{K}^{\rm st}\to-\frac{\epsilon^2+4\epsilon+2}{2(2+\epsilon)^2}\,,
\label{eq:K_hp}
\eea
and, with $\tilde r=r/a$,
\bea
P(\tilde r)=\sqrt{\frac{2(1+\epsilon)}{\pi}}\,\frac{\tilde r\;e^{-(1+\epsilon)(\tilde r^2-1)/2}}{\sqrt{\tilde r^2-1}}\,,\quad \tilde r>1\,.
\label{eq:P_hp}
\eea

\medskip

\noindent
\textbf{How robust is $-1/4$?~}
The kurtosis decreases monotonically from $-1/4$ ($\epsilon=0$) to the isotropic $-1/2$ ($\epsilon\to\infty$), so $\muperp k\lesssim\dr$, i.e. $\epsilon\lesssim1$, keeps the saturation value above $-7/18\simeq-0.389$ (substituting $\epsilon=1$ in $\mathcal{K}_{\rm st}$ in Eq.~\eqref{eq:K_hp}), while the minimum near $\chi^*$ reaches $\mathcal{K}^{{\rm st},*}\simeq-0.464$ at $\muperp=0$.
Reaching $-1/4$ itself requires strong anisotropy, because the $\mathcal{O}(\chi)$ correction is not small where $\epsilon$ is.
Expanding the exact result at fixed $\epsilon$,
\bea
\mathcal{K}^{\rm st}=-\frac{\epsilon^2+4\epsilon+2}{2(2+\epsilon)^2}
-\frac{6\epsilon^2+22\epsilon+13}{(1+\epsilon)(2+\epsilon)^2}\,\chi+\mathcal{O}(\chi^2),
\label{eq:Kcorr}
\eea
so $\mathcal{K}^{\rm st}\simeq-\tfrac14-\epsilon/4-13\chi/4$ for $\epsilon,\chi\ll1$.
Since $\epsilon=\lambda/\chi$, the intermediate shoulder is highest at
\bea
\chi_{\rm sh}=\sqrt{\lambda/13}\,,\qquad
\mathcal{K}^{\rm st}_{\rm sh}=-\frac14-\frac{\sqrt{13}}{2}\sqrt{\lambda}\,,
\label{eq:shoulder}
\eea
so $-1/4$ is approached only as $\sqrt{\lambda}$ and is resolved within $\delta$ for $\lambda\lesssim4\delta^2/13$, i.e. $\lambda\lesssim1.2\times10^{-4}$ for $\delta=0.02$.
The non-monotonic dependence of $\mathcal{K}^{\rm st}$ on $\chi$ survives for $\lambda<\lambda_c\simeq0.266$, with the minimum shifting to smaller $\chi$ as $\muperp$ grows ($\chi^*\simeq0.215$ at $\lambda=0.1$).
By contrast, the singular point of Eq.~\eqref{eq:rav_anisotropic}, generalized to $3\dr^2-\dr(\xi+\zeta)-\xi\zeta=0$, shifts to larger $\chi$ ($\chi\simeq0.442$ at $\lambda=0.1$), which confirms that it is unrelated to $\chi^*$.
For a slender body in a bulk Newtonian fluid, $\muperp/\mupara\simeq 1/2>0.266$~\cite{Doi1986}, so the non-monotonic kurtosis and the high-persistence weight outside the ring lie outside the hydrodynamic-rod regime of Fig.~\ref{fig4}; they require wheeled, gliding, or structured-media particles (Supplemental Material, Fig.~\ref{SM:fig:S2}).

\clearpage
\onecolumngrid
\setcounter{equation}{0}
\setcounter{figure}{0}
\setcounter{table}{0}
\setcounter{section}{0}
\setcounter{subsection}{0}
\renewcommand{\theequation}{S\arabic{equation}}
\renewcommand{\thefigure}{S\arabic{figure}}
\renewcommand{\thetable}{S\arabic{table}}
\renewcommand{\thesection}{S\Roman{section}}
\renewcommand{\theHequation}{S\arabic{equation}}
\renewcommand{\theHfigure}{S\arabic{figure}}
\renewcommand{\theHtable}{S\arabic{table}}
\renewcommand{\theHsection}{S\Roman{section}}

\begin{center}
{\large\bfseries Supplemental Material for:\\[2pt] Mobility Anisotropy Reshapes Self-Propelled Motion\par}
\vspace{0.8em}
Amir Shee and P. S. Pal
\end{center}
\vspace{0.5em}

\addtocontents{toc}{\protect\fi}
\tableofcontents

\medskip

\noindent
This Supplemental Material presents detailed moment calculations for both isotropic and anisotropic mobility, including exact Laplace-space derivations, time-dependent results, and the steady-state fourth moment and excess kurtosis (non-Gaussian parameter).
Section~\ref{SM:sec:general} extends the exact results to arbitrary transverse mobility $0\le\muperp\le\mupara$ and quantifies the accuracy of the perfectly anisotropic limit.

\section{Moments for Isotropic Mobility (\texorpdfstring{$\mupara=\muperp=\mu$}{mu-parallel = mu-perp = mu})}
\label{SM:sec:iso}

\noindent
We present the moments of the self-propelled particle with isotropic mobility
for completeness; the detailed calculations were first given in
Ref.~\cite{Chaudhuri2021}, and the closed-form expressions are also presented in the End Matter.
These results serve as the baseline against which the impact of
mobility anisotropy is assessed.

\medskip

\noindent
{\bf Fokker-Planck equation:~}
The probability distribution $P(\rv, \uv, t)$ of position $\rv$ and orientation $\uv$ of a particle governed by the equations of motion in main text follows the Fokker-Planck equation
\bea
\p_t P(\rv, \uv, t) &=& \dr \nabla_\uv^2 P - v_0\, \uv\cdot \nabla P - \mu (\nabla \cdot \fv) P - \mu (\fv \cdot \nabla)  P\,,
\label{SM:eq:F-P_iso}
\eea
where $\nabla,\nabla_\uv$ are gradient in position and orientation space, respectively.

\medskip

\noindent
Utilizing the Laplace transform $\tilde P(\rv, \uv, s) = \int_0^\infty dt\, e^{-s t}\, P(\rv, \uv, t) $, the Fokker-Planck Eq.~(\ref{SM:eq:F-P_iso}) is recast in the form
\bea
s \tilde P(\rv,\uv,s) &=& P(\rv,\uv,0)+ \dr \nabla_\uv^2 \tilde P - v_0 \uv\cdot\nabla\tilde P - \mu (\nabla \cdot \fv) \tilde P- \mu (\fv \cdot \nabla) \tilde P\,.
\label{SM:eq:Laplace-F-P_iso}
\eea
{\bf Moments generator equation:~}
We define the Laplace-space mean of an observable $\la \psi \ra_s = \int d\rv \, d\uv\, \psi(\rv, \uv ) \tilde P(\rv, \uv, s)$.
Multiplying Eq.~(\ref{SM:eq:Laplace-F-P_iso}) by $\psi(\rv, \uv)$ and integrating over all $(\rv, \uv)$,  we obtain
\bea
 s \la \psi \ra_s &=&  \la \psi \ra_0 +\dr \la \nabla_\uv^2 \psi \ra_s + v_0\la\, \uv\cdot \nabla \psi \ra_s + \mu \la \fv \cdot \nabla \psi \ra_s,
\label{SM:eq:moment_iso}
\eea
where the initial condition gives $\la \psi \ra_0 = \int d\rv \, d\uv\,  \psi(\rv, \uv) P(\rv, \uv, 0)$.
Without any loss of generality, we take the initial condition to be $P(\rv, \uv, 0) = \delta(\rv-\rv_0) \delta(\uv - \uv_0)$, where $\rv_0$ and $\uv_0$ are the initial position and orientation, respectively.
We set the initial position of the particle at the origin, $(x,y)=(0,0)$, with its initial orientation along the $x$-axis.
The external force is taken to be a harmonic restoring force, $\fv=-k\rv$.
Substituting this force into Eq.~\eqref{SM:eq:moment_iso}, we obtain closed expressions for the time evolution of the corresponding moments.

\medskip

\noindent
{\bf Mean orientation:}~
Setting $\psi=\uv$ gives $\la\uv\ra_s=\uv_0/(s+\dr)$,
so that $\la\uv(t)\ra=e^{-\dr t}\uv_0$ and the
well-known orientation autocorrelation follows as
$\la\uv(t)\cdot\uv(0)\ra=e^{-\dr t}$.

\medskip

\noindent
{\bf Mean displacement:~}
Substituting $\psi=\rv$ in Eq.~\eqref{SM:eq:moment_iso} gives
\bea
\la\rv\ra_s = \frac{v_0\la\uv\ra_s}{s+\mu k}\,,
\eea
and inverse Laplace transformation yields
the mean displacement
\bea
\la\rv\ra &=& -\f{v_0\uv_0\left(e^{-\dr t} - e^{-\mu k t}\right)}{\dr-\mu k}\,.
\label{SM:eq:rav_iso}
\eea
The singularity at $\dr=\mu k$ is removable, with $\la\rv\ra=v_0\uv_0\,t\,e^{-\dr t}$ at this point.
There is a single extremum, a maximum, at
\bea
t^{*}_{\rm iso} &=& \frac{\ln{(\mu k/\dr)}}{\mu k - \dr}
\eea
with $\mu k \neq \dr$, and its height is
\bea
\la\rv\ra_{\rm max} &=& \frac{v_0}{\dr} \left(\frac{\mu k}{\dr}\right)^{-\mu k /(\mu k -\dr)}
\;=\;\frac{v_0}{\mu k} \left(\frac{\mu k}{\dr}\right)^{-\dr /(\mu k -\dr)}\,.
\eea
For isotropic mobility, the peak time
$t^*_{\rm iso}$ decreases monotonically as $D_r$ increases, since stronger rotational diffusion reduces orientation persistence.
In the high-persistence limit the height tends to $v_0/\mu k$ from below, so the isotropic mean displacement never overshoots the force-balance value (contrast Sec.~\ref{SM:sec:general}).
At small times Eq.~\eqref{SM:eq:rav_iso} expands as
\bea
\la \rv \ra
&=& v_0 \uv_0
t- v_0 \uv_0\frac{\dr+\mu k}{2}\,t^{2} + v_0\,\uv_0 \frac{\dr^{2}+\dr\,\mu k+\mu^{2}k^{2}}{6}\,t^{3} + \mathcal{O}(t^{4}).
\label{SM:eq:rav_iso_small}
\eea
The leading term is ballistic; the $t^2$ correction encodes
independent suppression by rotational diffusion and the harmonic
restoring force.

\medskip

\noindent
{\bf Strict confinement of the isotropic steady state:}~
For isotropic mobility and $\fv=-k\rv$ the radial speed obeys
\bea
\f{d|\rv|}{dt}=v_0\,(\uv\cdot\hat\rv)-\mu k\,|\rv|\le v_0-\mu k\,|\rv|\,,
\eea
so $|\rv(t)|\le v_0/(\mu k)$ for all $t$ and all $\dr$, given $\rv_0=\mathbf{0}$.
The scaled steady-state distribution is therefore strictly supported on $\tilde r\le1$, which is the bound observed in Fig.~\ref{fig4}(d--f) of the main text.
No such bound holds for $\muperp=0$, where the transverse coordinate is not restored.

\medskip

\noindent
{\bf Position-orientation cross-correlation:}~
Setting $\psi=\uv\cdot\rv$ in Eq.~\eqref{SM:eq:moment_iso} and using
$\nabla_{\uv}^2(\uv\cdot\rv)=-\uv\cdot\rv$,
$\uv\cdot\nabla(\uv\cdot\rv)=1$, and $\rv\cdot\nabla(\uv\cdot\rv)=\uv\cdot\rv$:
\bea
\la\uv\cdot\rv\ra_s = \frac{v_0}{s(s+\dr+\mu k)}\,,
\label{SM:eq:ur_Laplace_iso}
\eea
giving $\la\uv\cdot\rv\ra = [v_0/(\dr+\mu k)](1-e^{-(\dr+\mu k)t})$,
identical in form to the anisotropic result [Eq.~\eqref{eq:urav_anisotropic} of the End Matter]
under $\mupara\to\mu$.

\medskip

\noindent
{\bf Mean-squared displacement (MSD):}~
Replacing $\psi=\rv^2$ in Eq.~\eqref{SM:eq:moment_iso} with $\uv\cdot\nabla(\rv^2)=2\uv\cdot\rv$  and
$\rv\cdot\nabla(\rv^2)=2\rv^2$:
\bea
\la\rv^2\ra_s = \frac{2v_0\la\uv\cdot\rv\ra_s}{s+2\mu k}
= \frac{2v_0^2}{s(s+\dr+\mu k)(s+2\mu k)}\,.
\label{SM:eq:r2_Laplace_iso}
\eea
Inverse transformation gives
\bea
\la\rv^2\ra
= \frac{v_0^2}{\mu k(\dr+\mu k)}
- \frac{2v_0^2\,e^{-\mu k t}}{\dr-\mu k}
  \!\left[\frac{e^{-\mu k t}}{2\mu k}
         -\frac{e^{-\dr t}}{\dr+\mu k}\right]\!.
\label{SM:eq:r2_iso}
\eea
The small-time expansion reads
\bea
\la\rv^2\ra
&=& v_0^2 t^2
  - \frac{v_0^2}{3}(\dr+3\mu k)\,t^3+ \frac{v_0^2}{12}(\dr^2+4\dr\mu k+7\mu^2 k^2)\,t^4
  + \mathcal{O}(t^5)\,.
  \label{SM:eq:r2v_iso_st}
\eea
The steady-state MSD is
\bea
\la\rv^2\ra_{\rm st}
= \frac{v_0^2}{\mu k(\dr+\mu k)}\,,
\label{SM:eq:r2_iso_st}
\eea
and the frozen-orientation ($\dr\to 0$) steady-state value coincides with the quasi-steady value for the anisotropic case
$\la\rv^2\ra_{\rm qs}=v_0^2/(\mu k)^2$.

\medskip

\noindent
{\bf Components of mean-squared displacement (MSD):~}
Due to the persistence of motion, the fixing of initial
active orientation of the ABP leads to asymmetric displacements, characterized by $\la\rv_{\parallel}^2\ra$ and $\la\rv_{\perp}^2\ra$, where $\rv_{\parallel}=\rv\cdot\uv_0$ and $\rv_{\perp}\perp\uv_0$.
Without any loss of generality, we assume that the initial orientation of activity along $x$-direction $\uv_0=\hat{x}$.
Using $\psi=\rv_{\parallel}^2=x^2$ in Eq.~\eqref{SM:eq:moment_iso}, we get
\bea
\la\rv_{\parallel}^2\ra_s &=& \f{2 v_0}{s+2\mu k} \la x u_x \ra_s\,.
\eea
Proceeding further, we consider $\psi = x u_x$ in Eq.~\eqref{SM:eq:moment_iso}, using $\la\psi\ra_0=0$, $\nabla_{\uv}^2\psi=-x u_x$, $\uv\cdot\nabla\psi=u_x^2$, and $\rv\cdot\nabla\psi=x u_x$, we obtain
\bea
\la x u_x\ra_s &=& \f{v_0 \la u_x^2\ra_s}{s+D_r+\mu k}\,.
\eea
To complete, we need to calculate $\la u_x^2\ra_s$. Using $\la u_x^2\ra_0=1$, $\la\nabla_{\uv}^2 u_x^2\ra_s=-4\la u_x^2\ra_s +2 \la 1\ra_s$, and $\uv\cdot\nabla u_x^2 = 0$, we find
\bea
\la u_x^2 \ra_s &=& \f{s + 2 D_r}{s(s+4 D_r)}\,.
\eea
Finally,
\bea
\la\rv_{\parallel}^2\ra_s &=& \f{2 v_0^2 (s+2 D_r)}{s (s+2\mu k) (s+D_r+\mu k) (s+4 D_r)}\,.
\label{SM:eq:rpara2_iso}
\eea
Performing inverse Laplace transform one finds
\begin{align}
\la\rv_\parallel^2\ra
&= \frac{v_0^2}{2\mu k(\dr+\mu k)}
  +\frac{v_0^2\,e^{-2\mu k t}}{\mu k(\mu k-2\dr)}+\frac{2v_0^2\,e^{-(\dr+\mu k)t}}{(\dr+\mu k)(3\dr-\mu k)}
  +\frac{v_0^2\,e^{-4\dr t}}{2(2\dr-\mu k)(3\dr-\mu k)}\,.
  \label{SM:eq:x2avg_iso}
\end{align}
At very short times, $t \ll 1/D_r,\,1/(\mu k)$, the motion is effectively ballistic along the initial propulsion direction. Expanding for small times
\bea
\langle \rv_{\parallel}^2\rangle = v_0^2 t^{2}
- (D_r + \mu k)\,v_0^2\, t^{3}
+ \frac{1}{12} \left(11 D_r^2 + 10 D_r \mu k + 7 (\mu k)^2\right) v_0^2\, t^{4}
+ \mathcal{O}(t^{5})\,,
\eea
showing that the leading correction to the ballistic growth arises from both rotational diffusion and the restoring force of the trap. At long times, $t \gg 1/D_r,\,1/(\mu k)$, all exponential terms decay, and the MSD along $\hat{x}$ saturates to the steady-state plateau
\bea
\langle \rv_{\parallel}^2\rangle_{\mathrm{st}} = \frac{v_0^2}{2\mu k (D_r+\mu k)}.
\eea

\medskip

\noindent
The perpendicular component of MSD can be calculated as $\la\rv_{\perp}^2\ra = \la\rv^2\ra - \la \rv_{\parallel}^2\ra$.
For $t\ll 1/D_r,\,1/(\mu k)$ the perpendicular component has no ballistic
$t^2$ term and begins at order $t^3$.  Its small-time expansion is
\bea
\langle \rv_{\perp}^2\rangle
= \frac{2}{3}\,D_r v_0^2\,t^{3}
- \frac{D_r\bigl(5 D_r + 3\mu k\bigr)}{6}\,v_0^2\,t^{4}
+ \mathcal{O}(t^{5})\,.
\eea
The steady-state value
\bea
\big\langle \rv_{\perp}^2\big\rangle_{\rm st}
&=& \frac{v_0^2}{2\,\mu k (D_r+\mu k)}\,.
\eea

\medskip

\noindent{\bf Displacement fluctuations and its components:~}
The fluctuations of the particle displacement defined as $\la \delta \rv^{2}\ra
= \la \rv^{2}\ra
- \la \rv\ra^{2}$ becomes
\bea
\la \delta \rv^2 \ra
&=&
\frac{v_0^2}{\mu k(D_r+ \mu k)}
-\frac{2v_0^2\, e^{- \mu k t}}{D_r-\mu k}
\left[
\frac{e^{- \mu k t}}{2 \mu k}
-\frac{e^{-D_r t}}{D_r + \mu k}
\right]
-\frac{v_0^2\left(e^{-D_r t} - e^{-\mu k t}\right)^2}{(D_r-\mu k)^2}\,.
\eea
At small times, the expansion reads
\bea
\la \delta \rv^2\ra
= \frac{2}{3}D_r v_0^2\,t^3
-\frac{1}{2}D_r(D_r+\mu k)\,v_0^2\,t^4
+ \mathcal{O}(t^5)\,.
\eea
The $t^{4}$ correction contains both $D_r$ and $\mu k$, showing how the trap
begins to suppress fluctuations once the particle starts to feel the restoring
force.

\medskip
\noindent
Projecting along the initial orientation, $\langle \delta \rv_\parallel^{2}\rangle
= \langle \rv_\parallel^{2}\rangle
- \langle \rv_\parallel\rangle^{2}$,
and its small-time expansion is
\bea
\la \delta \rv_\parallel^2 \ra
=
\frac{1}{3}\,D_r^{2}v_0^2\,t^{4}
- \frac{D_r^{2} v_0^{2}\,\bigl(7D_r + 3\mu k\bigr)}{15}\,t^{5}
+ \mathcal{O}(t^{6})\,.
\eea
The leading term is independent of $\mu k$: at this order the longitudinal fluctuations arise from rotational diffusion alone.

\medskip

\noindent
{\bf Second moment of the longitudinal displacement:}~
Setting $\psi=(\uv\cdot\rv)^2$ in Eq.~\eqref{SM:eq:moment_iso} with
$\nabla_{\uv}^2(\uv\cdot\rv)^2=2\rv^2-4(\uv\cdot\rv)^2$,
$\uv\cdot\nabla(\uv\cdot\rv)^2=2(\uv\cdot\rv)$,
$\rv\cdot\nabla(\uv\cdot\rv)^2=2(\uv\cdot\rv)^2$:
\bea
\la(\uv\cdot\rv)^2\ra_s
= \frac{2\dr\la\rv^2\ra_s+2v_0\la\uv\cdot\rv\ra_s}{s+4\dr+2\mu k}\,.
\eea
Substituting Eqs.~\eqref{SM:eq:r2_Laplace_iso} and~\eqref{SM:eq:ur_Laplace_iso}:
\begin{align}
\la(\uv\cdot\rv)^2\ra_s
&= \frac{2v_0^2(s+2\dr+2\mu k)}
        {s(s+2\mu k)(s+\dr+\mu k)(s+4\dr+2\mu k)}\,.
\label{SM:eq:ur2_Laplace_iso}
\end{align}
Inverse transformation yields
\begin{align}
\la(\uv\cdot\rv)^2\ra
&= \frac{v_0^2}{\mu k(2\dr+\mu k)}
  +\frac{v_0^2\,e^{-2\mu k t}}{2\mu k(\mu k-\dr)}-\frac{2v_0^2\,e^{-(\dr+\mu k)t}}{(\mu k-\dr)(3\dr+\mu k)}+\frac{v_0^2\,e^{-(4\dr+2\mu k)t}}{2(2\dr+\mu k)(3\dr+\mu k)}\,,
\label{SM:eq:ur2_iso}
\end{align}
with steady-state value
$\la(\uv\cdot\rv)^2\ra_{\rm st}=v_0^2/[\mu k(2\dr+\mu k)]$.

\medskip

\noindent
{\bf Steady-state fourth moment and non-Gaussian parameter:}~
We compute $\la\rv^4\ra_{\rm st}$ and $\mathcal{K}^{\rm st}$
using the shorthand $\xi\equiv\mu k$ and $\Gam\equiv\dr+\mu k$.
Setting $\psi=\rv^2(\uv\cdot\rv)$ in Eq.~\eqref{SM:eq:moment_iso} with
$\nabla_{\uv}^2[\rv^2(\uv\cdot\rv)]=-\rv^2(\uv\cdot\rv)$,
$\uv\cdot\nabla[\rv^2(\uv\cdot\rv)]=2(\uv\cdot\rv)^2+\rv^2$, and
$\rv\cdot\nabla[\rv^2(\uv\cdot\rv)]=3\rv^2(\uv\cdot\rv)$:
\bea
(s+\dr+3\xi)\la\rv^2(\uv\cdot\rv)\ra_s
= 2v_0\la(\uv\cdot\rv)^2\ra_s+v_0\la\rv^2\ra_s\,.
\label{SM:eq:r2ur_Laplace_iso}
\eea
Combining the right-hand side using Eqs.~\eqref{SM:eq:r2_Laplace_iso}
and~\eqref{SM:eq:ur2_Laplace_iso}:
\bea
\la\rv^2(\uv\cdot\rv)\ra_s
= \frac{2v_0^3(3s+8\dr+6\xi)}
       {s(s+\dr+3\xi)(s+\Gam)(s+2\xi)(s+4\dr+2\xi)}\,.
\label{SM:eq:r2ur_Laplace_iso_final}
\eea
Setting $\psi=\rv^4$ in Eq.~\eqref{SM:eq:moment_iso} with $\uv\cdot\nabla(\rv^4)=4\rv^2(\uv\cdot\rv)$
and $\rv\cdot\nabla(\rv^4)=4\rv^4$:
\bea
(s+4\xi)\la\rv^4\ra_s = 4v_0\la\rv^2(\uv\cdot\rv)\ra_s\,,
\eea
we obtain
\bea
\la\rv^4\ra_s
= \frac{8v_0^4(3s+8\dr+6\xi)}
       {s(s+4\xi)(s+\dr+3\xi)(s+\Gam)(s+2\xi)(s+4\dr+2\xi)}\,.
\label{SM:eq:r4_Laplace_iso}
\eea
Evaluating $\lim_{s\to 0}s\la\rv^4\ra_s$:
\bea
\la\rv^4\ra_{\rm st}
= \frac{v_0^4(4\dr+3\mu k)}
       {(\mu k)^2(\dr+3\mu k)(\dr+\mu k)(2\dr+\mu k)}\,.
\label{SM:eq:r4_iso_st}
\eea
The non-Gaussian parameter $\mathcal{K}^{\rm st}
=\la\rv^4\ra_{\rm st}/(2\la\rv^2\ra_{\rm st}^2)-1$ evaluates to
\bea
\mathcal{K}^{\rm st}
= -\,\frac{\mu k(7\dr+3\mu k)}
          {2(2\dr+\mu k)(\dr+3\mu k)}\,.
\label{SM:eq:alpha2_iso}
\eea
This is strictly non-positive, confirming a sub-Gaussian
steady state. Limiting behaviors:
\bea
\mathcal{K}^{\rm st}
&\xrightarrow{\;\dr/\xi\to 0\;}&
-\frac{1}{2}\,,
\label{SM:eq:alpha2_iso_low_Dr}\\[4pt]
\mathcal{K}^{\rm st}
&\xrightarrow{\;\xi/\dr\to 0\;}&
-\frac{7}{4}\,\frac{\mu k}{\dr}+\mathcal{O}\!\left(\frac{\mu k}{\dr}\right)^{\!2}
\to 0^-\,.
\label{SM:eq:alpha2_iso_high_Dr}
\eea
The activity-dominated value $-1/2$ reflects a
delta-ring steady-state distribution of radius $v_0/(\mu k)$,
obtained by averaging over all initial orientations
at frozen $D_r=0$.
For isotropic mobility the restoring term, $-\xi\,\rv\cdot\nabla\psi$, is the Euler operator and only rescales each monomial $r^{2m}(\uv\cdot\rv)^n$, so every relaxation rate is a linear combination of $\dr$ and $\xi$ and all denominators factorize into linear terms.
For anisotropic mobility the restoring term couples $r^{2m}(\uv\cdot\rv)^n$ to $r^{2m-2}(\uv\cdot\rv)^{n+2}$ of the same degree; the hierarchy still closes order by order (Sec.~\ref{SM:sec:general}), but the rates become roots of higher-order polynomials.

\section{Moments for Anisotropic Mobility
         (\texorpdfstring{$\mupara\neq 0$, $\muperp=0$}{mu-parallel != 0, mu-perp = 0})}
\label{SM:sec:aniso}

\noindent
{\bf Fokker-Planck equation:~}
The probability distribution $P(\rv, \uv, t)$ of the SPP follows
\bea
\p_t P &=& \dr\nabla_{\uv}^2 P
         - v_0\,\uv\cdot\nabla P
         - \mupara\nabla\cdot(\uv\uv^T\fv P)\,,
\label{SM:eq:FP_aniso}
\eea
where $\nabla,\nabla_\uv$ are gradient in spatial and orientation space, respectively.

\medskip

\noindent
Utilizing the Laplace transform $\tilde P(\rv, \uv, s) = \int_0^\infty dt\, e^{-s t}\, P(\rv, \uv, t) $, the Fokker-Planck Eq.~(\ref{SM:eq:FP_aniso}) is recast in the form
\bea
s \tilde P(\rv,\uv,s) &=& P(\rv,\uv,0)+ \dr \nabla_\uv^2 \tilde P - v_0 \uv \cdot \nabla \tilde P - \mupara\, (\uv\cdot\nabla)\big[(\uv\cdot\fv)\tilde P\big]\,.
\label{SM:eq:Laplace-F-P}
\eea
{\bf Moments generator equation:~}
We define the mean of an observable $\la \psi \ra_s = \int d\rv \, d\uv\, \psi(\rv, \uv ) \tilde P(\rv, \uv, s)$. Multiplying Eq.~(\ref{SM:eq:Laplace-F-P}) by $\psi(\rv, \uv)$ and integrating over all possible $(\rv, \uv)$, we find
\bea
 s \la \psi \ra_s &=&  \la \psi \ra_0 +\dr \la \nabla_\uv^2 \psi \ra_s + v_0\la\, \uv\cdot \nabla \psi \ra_s + \mupara \la \uv \uv^{T} \fv \cdot \nabla \psi \ra_s,
\label{SM:eq:moment}
\eea
where the initial condition sets $\la \psi \ra_0 = \int d\rv \, d\uv\,  \psi(\rv, \uv) P(\rv, \uv, 0)$. Without any loss of generality, we consider the initial condition $P(\rv, \uv, 0) = \delta(\rv-\rv_0) \delta(\uv - \uv_0)$, where $\rv_0$ and $\uv_0$ are the initial position and orientation, respectively. We set the initial position of the particle at the origin, $(x,y)=(0,0)$, with its initial orientation along the $x$-axis.
We consider the external force as a harmonic restoring force, $\fv=-k\rv$. Substituting this force into Eq.~(\ref{SM:eq:moment}), we derive closed expressions for the time evolution of the corresponding moments.

\medskip

\noindent
{\bf Mean orientation:}~
Identical to the isotropic case:
$\la\uv\ra_s=\uv_0/(s+\dr)$,
$\la\uv\ra=e^{-\dr t}\uv_0$.

\medskip

\noindent
{\bf Mean displacement:~}
Using $\psi = \rv$ in the Eq.~(\ref{SM:eq:moment}) with $\la\psi\ra_0=0$, $\nabla_{\uv}^2 \psi  = 0$, $\uv\cdot\nabla\psi=\uv$, we obtain
\bea
\la\rv\ra_s &=& \frac{v_0 \la\uv\ra_s-\mupara k\la (\uv\cdot\rv)\uv
\ra_s}{s}\,.
\label{SM:eq:rav_Laplace_harmonic}
\eea
To calculate $\la(\uv\cdot\rv)\uv\ra_s$, we use $\psi = (\uv\cdot\rv)\uv$ in Eq.~(\ref{SM:eq:moment}) with $\la\psi\ra_0=0$, $\nabla_{\uv}^2 \psi  = 2 (\rv-2(\uv\cdot\rv)\uv)$, $\uv\cdot\nabla\psi=\uv$, and get
\bea
\la(\uv\cdot\rv)\uv\ra_s = \frac{2\dr \la \rv \ra_s + v_0 \la\uv\ra_s}{(s+4\dr+\mupara k)}\,.
\eea
Substituting $\la(\uv\cdot\rv)\uv\ra_s$ and $\la\uv\ra_s$ back into Eq.~\eqref{SM:eq:rav_Laplace_harmonic} leads to the closed-form expression
\bea
\la\rv\ra_s &=& \f{ (s+4\dr)v_0 \uv_0 }{(s+D_r)[s(s+4\dr)+(s+2\dr)\mupara k]}\,.
\label{SM:eq:rav_Laplace_harmonic_final}
\eea

\medskip

\noindent
Inverse Laplace transform of Eq.~\eqref{SM:eq:rav_Laplace_harmonic_final} leads to Eq.~\eqref{eq:rav_anisotropic} presented in the main text.
The small time expansion of $\la\rv\ra$ leads to
\bea
\la\rv\ra &=& v_0 t \uv_0 - \f{v_0 \uv_0}{2} (D_r + \mupara k) t^2 + \f{v_0 \uv_0}{6} \left[ D_r^2 + 3 D_r \mupara k + (\mupara k)^2 \right] t^3 + \mathcal{O}(t^4) .
\eea
Comparing with Eq.~\eqref{SM:eq:rav_iso_small}, the departure from isotropy features in the $\dr\mupara k$ cross-term which acquires coefficient $3$ rather than $1$.

\medskip

\noindent
{\bf Removable singularity at $\mupara k=3\dr$:~}
The residue of Eq.~\eqref{SM:eq:rav_Laplace_harmonic_final} at $s=-\dr$ is $3v_0\uv_0/(\mupara k-3\dr)$, which produces the prefactor $1/(3\dr-\mupara k)$ in Eq.~\eqref{eq:rav_anisotropic} of the main text.
At $\mupara k=3\dr$ the quadratic in the denominator becomes $(s+\dr)(s+6\dr)$, so the slow root coincides with the orientational pole and $s=-\dr$ becomes a double pole.
The bracket in Eq.~\eqref{eq:rav_anisotropic} vanishes at the same point, and the inverse transform gives the regular result
\bea
\la\rv\ra\big|_{\mupara k=3\dr}
= v_0\uv_0\left[\frac{3}{5}\,t\,e^{-\dr t}+\frac{2}{25\dr}\left(e^{-\dr t}-e^{-6\dr t}\right)\right],
\label{SM:eq:rav_resonant}
\eea
which satisfies $\la\rv\ra\simeq v_0\uv_0 t$ at small times.
Equivalently, at $\mupara k=3\dr$ the rate $\Gam=\dr+\mupara k$ equals $4\dr$, the relaxation rate of the nematic tensor $\uv\uv^T-\mathbb{I}/2$ through which $\mupara\uv\uv^T$ couples translation and orientation.
The same condition sets the amplitude of the $e^{-\Gam t}$ mode in Eq.~\eqref{eq:r2av_anisotropic} of the main text to zero, since the numerator $s+4\dr$ of Eq.~\eqref{SM:eq:r2av_anisotropic_Laplace_final} vanishes at $s=-\Gam$.
The isotropic mean, Eq.~\eqref{SM:eq:rav_iso}, has the analogous removable singularity at $\mu k=\dr$.

\medskip

\noindent
{\bf Position-orientation cross correlation:~}
In order to calculate the cross correlation between orientation and position, we use $\psi=\uv\cdot\rv$ in Eq.~(\ref{SM:eq:moment}), with $\la\psi\ra_0=0$, $\nabla_{\uv}^2 \psi  = -\uv\cdot\rv$, $\uv\cdot\nabla\psi=1$, which gives
\bea
\la\uv\cdot\rv\ra_s = \f{v_0 \la 1\ra_s}{s+\dr + \mupara k}\,,
\label{SM:eq:uravg_Laplace}
\eea
where $\la 1 \ra_s=1/s$.
Inverse Laplace transformation of Eq.~\eqref{SM:eq:uravg_Laplace} leads to End Matter Eq.~\eqref{eq:urav_anisotropic}.

\medskip

\noindent
{\bf Mean-squared displacement (MSD):~}
To calculate $\la\rv^{2}\ra_s$ we use Eq.~(\ref{SM:eq:moment}) with $\psi=\rv^2$, $\la\psi\ra_0=0$, $\nabla_{\uv}^2\psi = 0$, and $\uv\cdot\nabla\psi=2(\uv\cdot\rv)$, which gives
\bea
s\la\rv^{2}\ra_s &=& 2 v_0 \la\uv\cdot\rv\ra_s - 2 \mupara k \la(\uv\cdot\rv)^2\ra_s\,,
\label{SM:eq:r2avg-Laplace-1}
\eea
where $\la\uv\cdot\rv\ra_s$ is given by Eq.~(\ref{SM:eq:uravg_Laplace}). To proceed further we calculate $\la(\uv\cdot\rv)^2\ra_s$ using $\psi=(\uv\cdot\rv)^2$, with $\la\psi\ra_0=0$, $\nabla_{\uv}^2\psi = 2 \rv^2 - 4 (\uv\cdot\rv)^2$, and $\uv\cdot\nabla\psi=2(\uv\cdot\rv)$:
\bea
\la(\uv\cdot\rv)^2\ra_s = \f{2\dr\la\rv^{2}\ra_s+2 v_0\la\uv\cdot\rv\ra_s}{s+4\dr+2\mupara k}\,.
\label{SM:eq:ur2av_anisotropic}
\eea
Substituting $\la(\uv\cdot\rv)^2\ra_s$ in Eq.~(\ref{SM:eq:r2avg-Laplace-1}), we get
\bea
\la\rv^{2}\ra_s = \f{(s+4\dr)2 v_0 \la\uv\cdot\rv\ra_s}{s(s+4\dr)+2(s+2\dr)\mupara k}\,.
\label{SM:eq:r2av-Laplace}
\eea
The final expression of MSD in Laplace space reads
\bea
\la\rv^{2}\ra_s &=& \f{2 v_0^2 (s+4 D_r)}{s(s+D_r +\mupara k) \left[s^2 + s (4D_r + 2\mupara k) + 4 \mupara k D_r\right]}\,.
\label{SM:eq:r2av_anisotropic_Laplace_final}
\eea
Inverse Laplace transform of Eq.~\eqref{SM:eq:r2av_anisotropic_Laplace_final} leads to Eq.~\eqref{eq:r2av_anisotropic} presented in the main text. In the small time limit, the MSD reads
\begin{align}
\la\rv^{2}\ra &\simeq v_0^2 t^2 - \f{v_0^2}{3} (\dr + 3 \mupara k)  t^3 + \f{v_0^2}{12} (\dr +\mupara k) (\dr + 7 \mupara k)  t^4+\mathcal{O}(t^5)\,,
\label{SM:eq:r2av_anisotropic_small_time}
\end{align}
where the cross-term coefficient of $D_r$ and $\mupara k$ at $\mathcal{O}(t^4)$ doubles ($4\dr\mu k\to 8\dr\mupara k$) relative to the isotropic case.
The steady-state MSD is
\bea
\la\rv^2\ra_{\rm st}
= \frac{2v_0^2}{\mupara k(\dr+\mupara k)}\,,
\label{SM:eq:r2_aniso_st}
\eea
a factor of two larger than the isotropic result (Eq.~\eqref{SM:eq:r2_iso_st}) at the same parameters, because transverse displacement is not directly restored
by the confinement.
In the frozen-orientation limit $\dr\to 0$,
$\la\rv^2\ra\big|_{\dr\to 0}=[v_0^2/(\mupara k)^2](1-e^{-\mupara k t})^2$,
giving the quasi-steady plateau
$\la\rv^2\ra_{\rm qs}=v_0^2/(\mupara k)^2$,
identical to the steady state value of the isotropic case.

\medskip

\noindent

\begin{figure}[!t]
\begin{center}
\includegraphics[width=\linewidth]{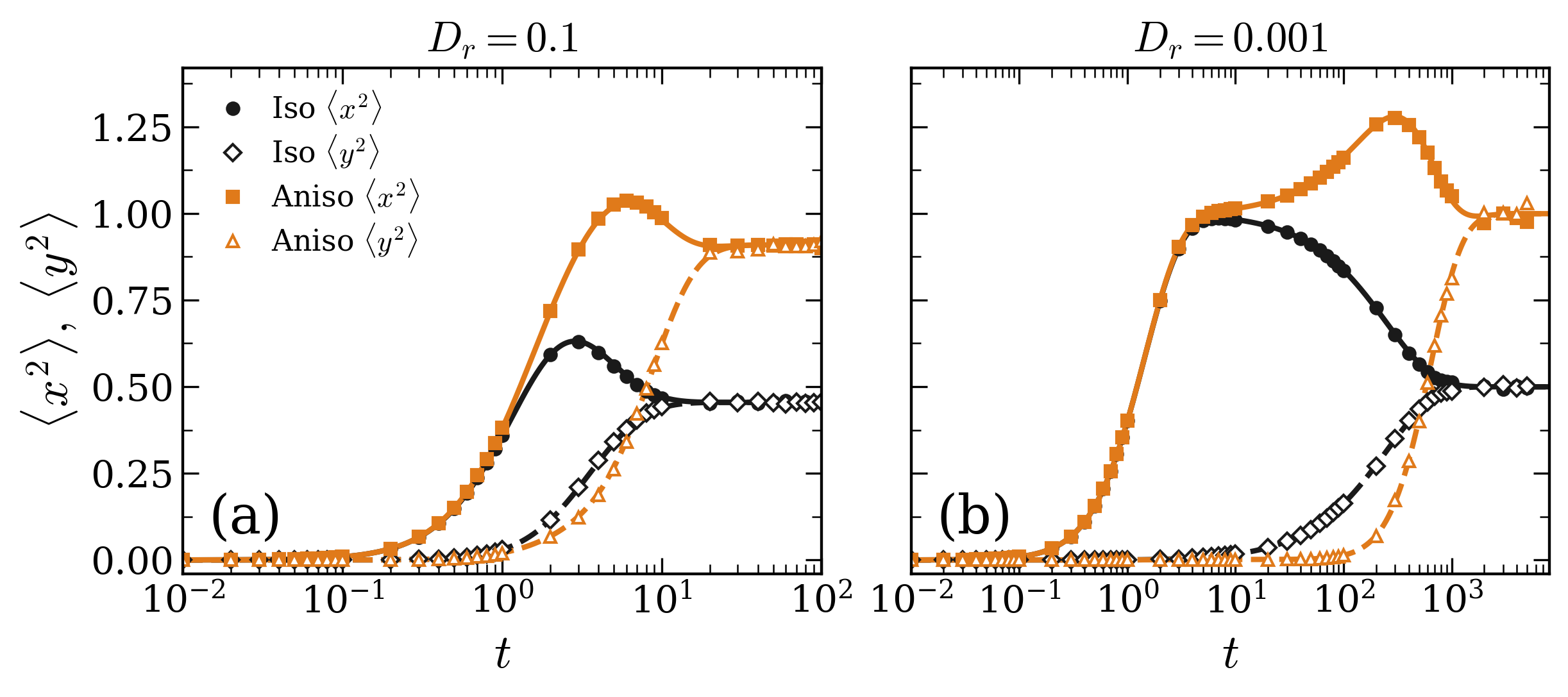}
\caption{
Components of the mean-squared displacement (MSD), $\la x^{2} \ra$ (solid lines) and $\la y^{2} \ra$ (dashed lines), for isotropic (black) and anisotropic (orange) mobility.
Lines are exact results, Eq.~(\ref{SM:eq:x2avg_iso}) for isotropic mobility and Eqs.~(\ref{SM:eq:n2sector}) and (\ref{SM:eq:x2y2}) for anisotropic mobility.
Symbols are simulation results averaged over $10^4$ active particles.
Parameters: $v_0=1$, $\mu=\mupara=1$, and $k=1$, with initial position $\rv=0$ and initial orientation along the $x$-axis.
}
\label{SM:app_fig1}
\end{center}
\end{figure}

\textbf{Components of mean-squared displacement (MSD):} For isotropic mobility, setting $\psi=x^2$ in Eq.~\eqref{SM:eq:moment_iso} gives the force term $\mu(-k\mathbf{r})\cdot\nabla(x^2)=-2\mu k x^2$, so the equation for $\langle x^2\rangle$ closes immediately. For anisotropic mobility the corresponding term,
\begin{equation}
\mu_\parallel k\,(\uv\cdot\mathbf{r})(\uv\cdot\nabla x^2)=2\mu_\parallel k\,(x^2u_x^2+xy\,u_xu_y),
\label{SM:eq:forcex2}
\end{equation}
introduces mixed position-orientation correlators. In the laboratory-frame basis $x^2\cos n\theta$ it couples each angular mode $\cos n\theta$ to $\cos[(n\pm2)\theta]$. In terms of the body-frame coordinates $\ell=\uv\cdot\mathbf{r}$ and $c=\uv_\perp\cdot\mathbf{r}$ of Sec.~\ref{SM:sec:general}, $z=x+iy=(\ell+ic)\,e^{i\theta}$. A joint rotation of $\mathbf{r}$ and $\uv$ by an angle $\varphi$ leaves $(\ell,c)$ unchanged and shifts $\theta\to\theta+\varphi$, so a moment $\langle f(\ell,c)\,e^{in\theta}\rangle$ acquires the factor $e^{in\varphi}$. Since the dynamics is invariant under such rotations, moments with different $n$ do not couple, and the moments $A^{(n)}_{pq}\equiv\langle \ell^{p}c^{q}e^{in\theta}\rangle$ close at each order $p+q$ and for each $n$, for arbitrary $\mu_\perp$. Since
\begin{equation}
x^2=\tfrac12\,\mathbf{r}^2+\tfrac12\,\mathrm{Re}\big[(\ell+ic)^2e^{2i\theta}\big],
\label{SM:eq:x2decomp}
\end{equation}
$\langle x^2\rangle$ requires, besides $\langle\mathbf{r}^2\rangle$, only the sector $n=2$ up to second order. Acting with the generator, Eq.~\eqref{SM:eq:generator_bf}, supplemented by the orientational terms $D_r\partial_\theta^2+2D_r(c\,\partial_\ell-\ell\,\partial_c)\partial_\theta$, and using $\mathbf{r}_0=0$ and $\theta_0=0$, we obtain in Laplace space, with $\xi=\mu_\parallel k$ and $\zeta=\mu_\perp k$,
\begin{align}
(s+4D_r)A_{00}&=1, \nonumber\\
(s+5D_r+\xi)A_{10}&=v_0A_{00}+4iD_rA_{01}, \qquad (s+5D_r+\zeta)A_{01}=-4iD_rA_{10}, \nonumber\\
(s+6D_r+2\xi)A_{20}&=2v_0A_{10}+2D_rA_{02}+8iD_rA_{11}, \nonumber\\
(s+8D_r+\xi+\zeta)A_{11}&=v_0A_{01}+4iD_r(A_{02}-A_{20}), \nonumber\\
(s+6D_r+2\zeta)A_{02}&=2D_rA_{20}-8iD_rA_{11},
\label{SM:eq:n2sector}
\end{align}
where $A_{pq}\equiv A^{(2)}_{pq}$. The Cartesian components follow as
\begin{equation}
\langle x^2\rangle_s=\tfrac12\langle\mathbf{r}^2\rangle_s+\tfrac12\big(A_{20}-A_{02}+2iA_{11}\big),\qquad \langle y^2\rangle_s=\langle\mathbf{r}^2\rangle_s-\langle x^2\rangle_s ,
\label{SM:eq:x2y2}
\end{equation}
where the combination in brackets is real. For $\zeta=\xi$ this reproduces Eq.~\eqref{SM:eq:rpara2_iso}. For $\zeta=0$ the $n=2$ sector relaxes at the rate $4D_r$ and at the rates $-s_i$, where $s_i$ are the zeros of $s^2+(10D_r+\xi)s+9D_r^2+5D_r\xi$ and of $s^3+(20D_r+3\xi)s^2+2(32D_r^2+20D_r\xi+\xi^2)s+4D_r\xi(16D_r+3\xi)$. All $n=2$ moments decay, so $\langle x^2\rangle_{\rm st}=\langle y^2\rangle_{\rm st}=\langle\mathbf{r}^2\rangle_{\rm st}/2$. The time dependence follows by numerical inversion and agrees with simulations in Fig.~\ref{SM:app_fig1}.

\medskip

\noindent
{\bf Second moment of the longitudinal displacement $\la (\uv\cdot\rv)^2 \ra$:~}
Substituting $\la\uv\cdot\rv\ra_s$ (Eq.~\eqref{SM:eq:uravg_Laplace}) and $\la\rv^2\ra_s$ (Eq.~\eqref{SM:eq:r2av_anisotropic_Laplace_final}) into Eq.~\eqref{SM:eq:ur2av_anisotropic}, we get
\begin{align}
\langle (\uv\cdot\rv)^2\rangle_s
&=
\frac{2v_0^2\,(s+2D_r)}
{s(s+D_r+\mupara k)[s^2+(4D_r+2\mu_\parallel k)s+4D_r\,\mu_\parallel k]}\,.
\label{SM:eq:ur2av_anisotropic_Laplace}
\end{align}

Inverse Laplace transform of the above equation leads to
\begin{align}
\langle (\uv\cdot\rv)^2\rangle
&=
\frac{v_0^2}{\mu_\parallel k\,(D_r+\mu_\parallel k)}
+\frac{2v_0^2\,(D_r-\mu_\parallel k) \,
e^{-(D_r+\mu_\parallel k)t}}{(D_r+\mu_\parallel k)\,\big(3D_r^2+(\mu_\parallel k)^2\big)}
\nonumber\\[4pt]
&\quad
-\frac{v_0^2\left(\sqrt{4D_r^2+(\mu_\parallel k)^2}-\mu_\parallel k\right)\,e^{-\left(2D_r+\mu_\parallel k-\sqrt{4D_r^2+(\mu_\parallel k)^2}\right)t}}
{\sqrt{4D_r^2+(\mu_\parallel k)^2}\;
\Big(2D_r+\mu_\parallel k-\sqrt{4D_r^2+(\mu_\parallel k)^2}\Big)\;
\Big(\sqrt{4D_r^2+(\mu_\parallel k)^2}-D_r\Big)}
\nonumber\\[4pt]
&\quad
+\frac{v_0^2\left(\sqrt{4D_r^2+(\mu_\parallel k)^2}+\mu_\parallel k\right)\,e^{-\left(2D_r+\mu_\parallel k+\sqrt{4D_r^2+(\mu_\parallel k)^2}\right)t}}
{\sqrt{4D_r^2+(\mu_\parallel k)^2}\;
\Big(2D_r+\mu_\parallel k+\sqrt{4D_r^2+(\mu_\parallel k)^2}\Big)\;
\Big(\sqrt{4D_r^2+(\mu_\parallel k)^2}+D_r\Big)}\,.
\label{SM:eq:ur2avg_anisotropic}
\end{align}
At short times, the second moment of the longitudinal displacement admits the expansion
\begin{equation}
\langle (\uv\cdot\rv)^2 \rangle
= v_0^2\,t^2
- v_0^2\,(D_r+\mupara k)\,t^3
+\frac{v_0^2}{12}\Big(11D_r^2+14D_r \mupara k+7 (\mupara k)^2\Big)\,t^4
+ \mathcal{O}(t^{5}) ,
\end{equation}
demonstrating the expected ballistic scaling $\la (\uv\cdot\rv)^2 \ra \sim v_0^2 t^2$ as $t \to 0$.
In the long-time limit, the steady-state value of the second moment
$\langle (\uv\cdot\rv)^2 \rangle$ is obtained by taking
$t\to\infty$ in Eq.~(\ref{SM:eq:ur2avg_anisotropic}), yielding the stationary
result
\begin{equation}
\label{SM:eq:ur2_squared_steady}
\big\langle (\uv\cdot\rv)^2 \big\rangle_{\mathrm{st}}
= \frac{v_0^{2}}{\mupara k (D_r+\mupara k)}.
\end{equation}
Physically, $\langle(\uv\cdot\rv)^2\rangle$ quantifies fluctuations of the particle
displacement along its instantaneous propulsion direction and thus provides a
direct measure of how persistent self-propulsion is converted into longitudinal
excursions against the restoring force of the harmonic trap.

In the long-time regime, the stationary value in
Eq.~(\ref{SM:eq:ur2_squared_steady}) results from a balance between three competing
effects: propulsion injects fluctuations along the body axis, rotational diffusion
randomizes the orientation and limits directional memory, while harmonic
confinement suppresses large displacements.
The transient rates $(D_r+\mupara k)$ and $(2D_r+\mupara k)\pm\sqrt{4\dr^2+(\mupara k)^2}$ reflect the coupling between orientational relaxation and
mechanical relaxation of the longitudinal coordinate.
In particular, stronger confinement or larger parallel mobility reduces the
steady-state longitudinal fluctuations, while decreasing $D_r$ enhances them by
increasing orientational persistence.
This highlights the intrinsically nonequilibrium nature of the steady state, in
which propulsion continuously feeds longitudinal fluctuations that are
dissipated by rotational noise and confinement.

\medskip

\noindent
{\bf Steady-state fourth moment and non-Gaussian parameter:}~
We use $\xi\equiv\mupara k$ and $\Gam\equiv\dr+\mupara k$.
The steady-state moment equations follow from
setting $\p_t\la\psi\ra=0$ (or equivalently, $s\la\psi\ra_s-\la\psi\ra_0=0$) in Eq.~\eqref{SM:eq:moment}:
\bea
0 = \dr\la\nabla_{\uv}^2\psi\ra_{\rm st}
  + v_0\la\uv\cdot\nabla\psi\ra_{\rm st}
  - \xi\la(\uv\cdot\rv)(\uv\cdot\nabla\psi)\ra_{\rm st}\,.
\label{SM:eq:ss_aniso}
\eea
The required lower-order inputs in steady state are
$\la\uv\cdot\rv\ra_{\rm st}=v_0/\Gam$,
$\la\rv^2\ra_{\rm st}=2v_0^2/(\xi\Gam)$, and
$\la(\uv\cdot\rv)^2\ra_{\rm st}=v_0^2/(\xi\Gam)$.

\smallskip
\noindent
\underline{\textit{Third-order moments:}}~
Defining $C_1\equiv\la\rv^2(\uv\cdot\rv)\ra_{\rm st}$ and
$C_2\equiv\la(\uv\cdot\rv)^3\ra_{\rm st}$, the observables
$\psi=\rv^2(\uv\cdot\rv)$ and $\psi=(\uv\cdot\rv)^3$ in
Eq.~\eqref{SM:eq:ss_aniso} yield
\bea
-\Gam\,C_1 + \frac{4v_0^3}{\xi\Gam} - 2\xi\,C_2 &=& 0\,,
\label{SM:eq:C1C2_A}\\
6\dr\,C_1 - 3(3\dr+\xi)\,C_2 + \frac{3v_0^3}{\xi\Gam} &=& 0\,.
\label{SM:eq:C1C2_B}
\eea
Solving these equations
\bea
C_1 &=& \frac{2v_0^3(6\dr+\xi)}
             {\xi\Gam(3\dr^2+8\dr\xi+\xi^2)}\,,
\label{SM:eq:C1}\\[4pt]
C_2 &=& \frac{v_0^3(9\dr+\xi)}
             {\xi\Gam(3\dr^2+8\dr\xi+\xi^2)}\,.
\label{SM:eq:C2}
\eea
The polynomial $3\dr^2+8\dr\xi+\xi^2$ first appears here as a direct
signature of the coupled orientation--translation relaxation imposed
by the anisotropic mobility tensor.
It is one third of the determinant of the $2\times2$ system~\eqref{SM:eq:C1C2_A}--\eqref{SM:eq:C1C2_B}, i.e., one third of the product of the two relaxation rates of the third-order sector [see $Q_3(0)$ below].

\smallskip
\noindent
\underline{\textit{Fourth-order moments:}}~
Defining $P\equiv\la\rv^4\ra_{\rm st}$,
$Q\equiv\la\rv^2(\uv\cdot\rv)^2\ra_{\rm st}$, and
$R\equiv\la(\uv\cdot\rv)^4\ra_{\rm st}$, the observables
$\psi=\rv^4$, $\psi=(\uv\cdot\rv)^4$, and
$\psi=\rv^2(\uv\cdot\rv)^2$ in Eq.~\eqref{SM:eq:ss_aniso} give,
respectively,
\bea
Q &=& \frac{v_0}{\xi}\,C_1
    = \frac{2v_0^4(6\dr+\xi)}
           {\xi^2\Gam(3\dr^2+8\dr\xi+\xi^2)}\,,
\label{SM:eq:Q}\\[6pt]
R &=& \frac{3\dr\,Q+v_0\,C_2}{4\dr+\xi}
    = \frac{v_0^4(36\dr^2+15\dr\xi+\xi^2)}
           {(4\dr+\xi)\xi^2\Gam(3\dr^2+8\dr\xi+\xi^2)}\,,
\label{SM:eq:R}\\[6pt]
\dr\,P &=& 2\dr\,Q+\xi(Q+R)-v_0(C_1+C_2)\,.
\label{SM:eq:P_eqn}
\eea
Substituting $C_1$, $C_2$, $Q$, and $R$ into
Eq.~\eqref{SM:eq:P_eqn} and collecting terms over the common
denominator $\xi^2(4\dr+\xi)\Gam(3\dr^2+8\dr\xi+\xi^2)$,
one finds
\bea
\la\rv^4\ra_{\rm st}
= \frac{2v_0^4(48\dr^2+20\dr\mupara k+3(\mupara k)^2)}
       {(\mupara k)^2(4\dr+\mupara k)(\dr+\mupara k)
        (3\dr^2+8\dr\mupara k+(\mupara k)^2)}\,.
\label{SM:eq:r4_aniso_st}
\eea
The non-Gaussian parameter
$\mathcal{K}^{\rm st}=\la\rv^4\ra_{\rm st}/(2\la\rv^2\ra_{\rm st}^2)-1$
evaluates to
\bea
\mathcal{K}^{\rm st}
= -\,\frac{\mupara k(72\dr^2+25\dr\mupara k+(\mupara k)^2)}
          {4(4\dr+\mupara k)(3\dr^2+8\dr\mupara k+(\mupara k)^2)}\,.
\label{SM:eq:alpha2_aniso}
\eea
This is strictly non-positive. Limiting behaviors:
\bea
\mathcal{K}^{\rm st}
&\xrightarrow{\;\dr/\xi\to 0\;}&
-\frac{1}{4}\,,
\label{SM:eq:alpha2_aniso_low_Dr}\\[4pt]
\mathcal{K}^{\rm st}
&\xrightarrow{\;\xi/\dr\to 0\;}&
-\frac{3}{2}\,\frac{\mupara k}{\dr}
+\mathcal{O}\!\left(\frac{\mupara k}{\dr}\right)^{\!2}
\to 0^-\,.
\label{SM:eq:alpha2_aniso_high_Dr}
\eea

\medskip

\noindent
{\bf Location of the kurtosis extremum:~}
With $\chi=\dr/\xi$, Eq.~\eqref{SM:eq:alpha2_aniso} depends on $\chi$ alone.
The numerator of $d\mathcal{K}^{\rm st}/d\chi$ is, up to a positive factor,
\bea
864\chi^4+600\chi^3+47\chi^2-74\chi-13\,,
\label{SM:eq:quartic}
\eea
which is irreducible over the rationals and has a single positive root,
\bea
\chi^*=0.331765\ldots\,,\qquad \mathcal{K}^{{\rm st},*}=-0.464287\ldots\,.
\label{SM:eq:chistar_value}
\eea
The quartic~\eqref{SM:eq:quartic} equals $4/9$ at $\chi=1/3$, so $\chi^*$ does not coincide with the removable singularity of Eq.~\eqref{SM:eq:rav_resonant}.
The two points also move in opposite directions when $\muperp>0$ (Sec.~\ref{SM:sec:general}).
This is consistent with the structure of the problem: $\mathcal{K}^{\rm st}$ involves only rotationally invariant moments, whereas the degeneracy occurs in the vector sector that governs $\la\rv\ra$.

\medskip

\noindent
{\bf Time-dependent fourth moment:~}
The equations used above for the steady state hold in Laplace space after replacing $0$ by $s\la\psi\ra_s-\la\psi\ra_0$, and they remain closed at every order.
Solving them gives
\bea
\la\rv^4\ra_s=\frac{8v_0^4\,N(s)}{s\,Q_1(s)\,Q_2(s)\,Q_3(s)\,Q_4(s)}\,,
\label{SM:eq:r4_Laplace_aniso}
\eea
with
\bea
N(s) &=& 3s^4+(95\dr+9\xi)s^3+(964\dr^2+182\dr\xi+6\xi^2)s^2
+(3680\dr^3+1148\dr^2\xi+60\dr\xi^2)s\nn\\
&&+\,4608\dr^4+1920\dr^3\xi+288\dr^2\xi^2\,,
\eea
and the characteristic polynomials of the first- to fourth-order sectors,
\bea
Q_1(s) &=& s+\Gam\,,\nn\\
Q_2(s) &=& s^2+(4\dr+2\xi)s+4\dr\xi\,,\nn\\
Q_3(s) &=& s^2+(10\dr+4\xi)s+3(3\dr^2+8\dr\xi+\xi^2)\,,\nn\\
Q_4(s) &=& s^3+(20\dr+6\xi)s^2+8(8\dr^2+10\dr\xi+\xi^2)s+32\dr\xi(4\dr+\xi)\,.
\label{SM:eq:Qn}
\eea
Equation~\eqref{SM:eq:r4_Laplace_aniso} satisfies $\la\rv^4\ra\simeq v_0^4t^4$ at small times and reproduces Eq.~\eqref{SM:eq:r4_aniso_st} for $s\to0$.
The steady-state denominator is thus built from $Q_1(0)=\Gam$, $Q_2(0)=4\dr\xi$, $Q_3(0)=3(3\dr^2+8\dr\xi+\xi^2)$, and $Q_4(0)=32\dr\xi(4\dr+\xi)$, which accounts for every factor in Eqs.~\eqref{SM:eq:r4_aniso_st} and~\eqref{SM:eq:alpha2_aniso}.
The time-dependent $\la\rv^4\ra$ and $\mathcal{K}(t)$ follow from Eq.~\eqref{SM:eq:r4_Laplace_aniso} by numerical inversion; a closed form requires the roots of the cubic $Q_4$.

\medskip

\noindent
{\bf Comparison of isotropic and anisotropic moments and non-Gaussian parameters:}~
Although both isotropic and anisotropic steady states are sub-Gaussian, they differ in two key
aspects:

(i)~\textit{Activity-dominated limit.}
The isotropic excess kurtosis saturates to $-1/2$, corresponding to
a delta-ring steady state at radius $v_0/(\mu k)$;
the anisotropic value $-1/4$ is less negative because
transverse confinement is absent and the distribution
remains more broadly spread [see Eq.~\eqref{SM:eq:P_hp}].

(ii)~\textit{Functional form.}
The denominator of the isotropic excess kurtosis (Eq.~\eqref{SM:eq:alpha2_iso})
contains only linear factors $(\dr+3\xi)$ and $(2\dr+\xi)$,
reflecting moment-hierarchy closure at each order without
coupling between rotational and translational relaxation.
The anisotropic denominator (Eq.~\eqref{SM:eq:alpha2_aniso})
contains the irreducible quadratic $3\dr^2+8\dr\xi+\xi^2$,
a structural fingerprint of the anisotropic mobility tensor
that first enters at third order (Eqs.~\eqref{SM:eq:C1}--\eqref{SM:eq:C2})
and propagates to the fourth moment through $Q_3(0)$.

\begin{table}[h]
\caption{Comparison of steady-state quantities for isotropic
($\mupara=\muperp=\mu$) and anisotropic ($\mupara\neq 0$,
$\muperp=0$) mobility. Here $\xi=\mu k$ (isotropic) or
$\xi=\mupara k$ (anisotropic).}
\begin{ruledtabular}
\begin{tabular}{lcc}
Quantity & Isotropic & Anisotropic \\[2pt]
\hline\\[-6pt]
$\la\rv^2\ra_{\rm st}$
  & $\dfrac{v_0^2}{\xi\Gam}$
  & $\dfrac{2v_0^2}{\xi\Gam}$ \\[10pt]
$\la(\uv\cdot\rv)^2\ra_{\rm st}$
  & $\dfrac{v_0^2}{\xi(2\dr+\xi)}$
  & $\dfrac{v_0^2}{\xi\Gam}$ \\[10pt]
$\la\rv^4\ra_{\rm st}$
  & $\dfrac{v_0^4(4\dr+3\xi)}{\xi^2(\dr+3\xi)\Gam(2\dr+\xi)}$
  & $\dfrac{2v_0^4(48\dr^2+20\dr\xi+3\xi^2)}{\xi^2(4\dr+\xi)\Gam(3\dr^2+8\dr\xi+\xi^2)}$\\[10pt]
$\mathcal{K}^{\rm st}$
  & $-\dfrac{\xi(7\dr+3\xi)}{2(2\dr+\xi)(\dr+3\xi)}$
  & $-\dfrac{\xi(72\dr^2+25\dr\xi+\xi^2)}{4(4\dr+\xi)(3\dr^2+8\dr\xi+\xi^2)}$\\[10pt]
$\mathcal{K}^{\rm st}\big|_{\dr\to 0}$
  & $-\dfrac{1}{2}$
  & $-\dfrac{1}{4}$\\[8pt]
$\mathcal{K}^{\rm st}\big|_{\dr=\xi}$
  & $-\dfrac{5}{12}\approx-0.42$
  & $-\dfrac{49}{120}\approx-0.41$\\[8pt]
$\mathcal{K}^{\rm st}\big|_{\xi\ll\dr}$
  & $-\dfrac{7\xi}{4\dr}$
  & $-\dfrac{3\xi}{2\dr}$\\[8pt]
\end{tabular}
\end{ruledtabular}
\label{SM:tab:comparison}
\end{table}

\begin{figure}[!t]
\begin{center}
\includegraphics[width=\linewidth]{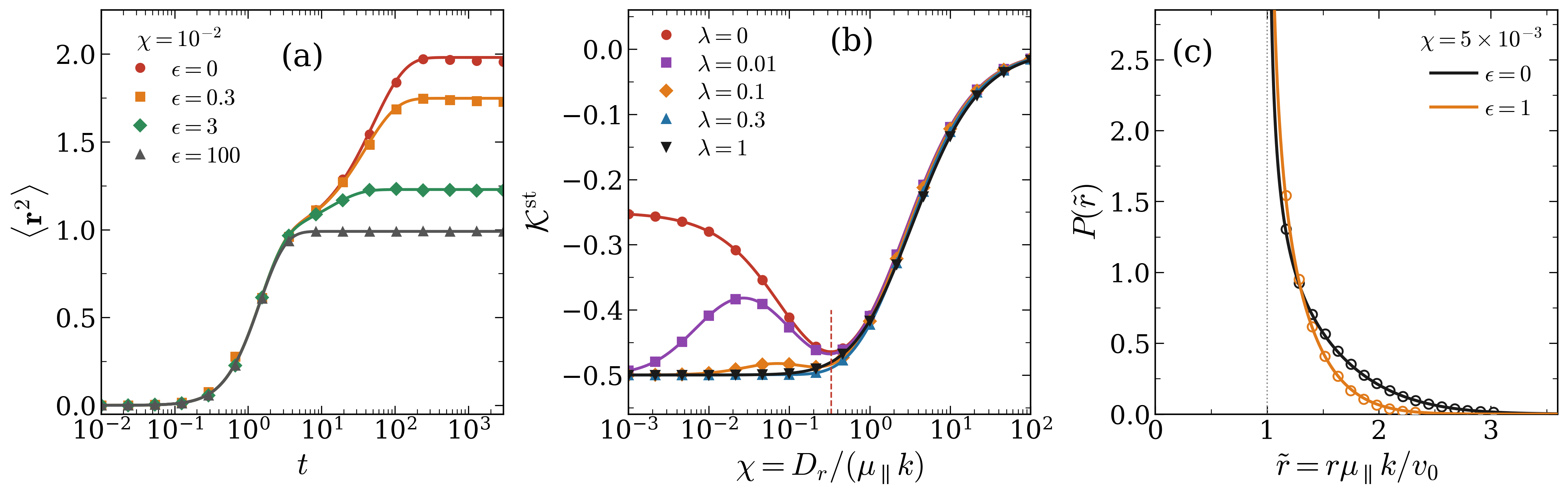}
\caption{
Finite transverse mobility ($v_0=\mupara=k=1$).
Lines are exact results; symbols are simulations.
(a) MSD at $\chi=10^{-2}$ for several $\epsilon=\muperp k/\dr$, obtained by numerical inversion of Eq.~\eqref{SM:eq:r2_Laplace_general}; dotted lines mark the long-time value of Eq.~\eqref{SM:eq:K_hp}, and the slow-sector form Eq.~\eqref{SM:eq:slow_msd} is indistinguishable from the exact curves on this scale.
(b) $\mathcal{K}^{\rm st}$ versus $\chi$ for several $\lambda=\muperp/\mupara$, from Eqs.~\eqref{SM:eq:r2_general_st} and~\eqref{SM:eq:r4_general_st}; the dashed line marks $\chi^*$ for $\lambda=0$.
The curves for $\lambda=10^{-4}$ and $10^{-5}$ are included to show the intermediate shoulder approaching $-1/4$, whose height is given by Eq.~\eqref{SM:eq:shoulder}.
(c) Steady-state radial distribution at $\chi=5\times10^{-3}$ compared with Eq.~\eqref{SM:eq:P_hp}.
}
\label{SM:fig:S2}
\end{center}
\end{figure}

\section{Arbitrary Transverse Mobility (\texorpdfstring{$0\le\muperp\le\mupara$}{0 <= mu-perp <= mu-parallel})}
\label{SM:sec:general}

\noindent
We now consider the general mobility tensor
\bea
\boldsymbol{\mu}=\mupara\,\uv\uv^T+\muperp\left(\mathbb{I}-\uv\uv^T\right),
\label{SM:eq:mu_general}
\eea
and write $\xi\equiv\mupara k$, $\zeta\equiv\muperp k$, $\Gam\equiv\dr+\xi$, $\lambda\equiv\muperp/\mupara$, $\chi\equiv\dr/\xi$, and $a\equiv v_0/\xi$.
The limits $\zeta=0$ and $\zeta=\xi$ recover Secs.~\ref{SM:sec:aniso} and~\ref{SM:sec:iso}.

\medskip

\noindent
{\bf Body-frame equations:~}
With $\fv=-k\rv$, the position obeys $\dot\rv=v_0\uv-\zeta\rv-(\xi-\zeta)(\uv\cdot\rv)\uv$.
We introduce the longitudinal and transverse coordinates $\ell\equiv\uv\cdot\rv$ and $c\equiv\uv_\perp\cdot\rv$, so that $r^2=\ell^2+c^2$.
In It\^o form, Eq.~\eqref{eom3:rot} of the main text reads $d\uv=-\dr\uv\,dt+\sqrt{2\dr}\,\uv_\perp dW$ and $d\uv_\perp=-\dr\uv_\perp dt-\sqrt{2\dr}\,\uv\,dW$.
Since $d\rv$ carries no noise, $d\ell=\uv\cdot d\rv+\rv\cdot d\uv$ and $dc=\uv_\perp\cdot d\rv+\rv\cdot d\uv_\perp$, which gives
\bea
d\ell &=& \big[v_0-(\dr+\xi)\,\ell\big]\,dt+\sqrt{2\dr}\;c\,dW\,,\nn\\
dc &=& -(\dr+\zeta)\,c\,dt-\sqrt{2\dr}\;\ell\,dW\,.
\label{SM:eq:bodyframe}
\eea
The corresponding backward generator acting on $f(\ell,c)$ is
\bea
\mathcal{L}f=\big[v_0-\Gam\ell\big]\p_\ell f-(\dr+\zeta)\,c\,\p_c f+\dr\left(c^2\p_\ell^2-2\ell c\,\p_\ell\p_c+\ell^2\p_c^2\right)f\,.
\label{SM:eq:generator_bf}
\eea
The drift is linear and the diffusion matrix is quadratic, so $\mathcal{L}$ maps polynomials of degree $n$ in $(\ell,c)$ onto polynomials of degree $\le n$.
Rotationally invariant moments are polynomials in $(\ell,c)$, so they close order by order for every $\muperp$.
Moments carrying a factor $e^{in\theta}$ close in the same way for each $n$, which gives the Cartesian components exactly (Sec.~\ref{SM:sec:aniso}).

\medskip

\noindent
{\bf Moment operator in the laboratory frame:~}
Equivalently, in terms of the basis $r^{2m}\ell^n$ used in Secs.~\ref{SM:sec:iso} and~\ref{SM:sec:aniso}, the generalization of Eq.~\eqref{SM:eq:moment} is
$s\la\psi\ra_s=\la\psi\ra_0+\la\mathcal{L}\psi\ra_s$ with
\bea
\mathcal{L}\big[r^{2m}\ell^n\big]
&=& \dr\left[n(n-1)\,r^{2m+2}\ell^{n-2}-n^2\,r^{2m}\ell^n\right]
+v_0\left[2m\,r^{2m-2}\ell^{n+1}+n\,r^{2m}\ell^{n-1}\right]\nn\\
&&-\,\zeta\,(2m+n)\,r^{2m}\ell^n
-(\xi-\zeta)\left[2m\,r^{2m-2}\ell^{n+2}+n\,r^{2m}\ell^n\right].
\label{SM:eq:master}
\eea
Rotational diffusion and the restoring force preserve the degree $2m+n$, while propulsion lowers it by one.
The moments of degree $N$ therefore form a closed linear system of dimension $\lfloor N/2\rfloor+1$, sourced by moments of lower degree.

\medskip

\noindent
{\bf Mean displacement and cross-correlation:~}
Setting $\psi=\ell$ in Eq.~\eqref{SM:eq:master} gives $\la\uv\cdot\rv\ra_s=v_0/[s(s+\Gam)]$, independent of $\muperp$, as stated in the End Matter.
For the vector moment, $\psi=\rv$ and $\psi=\ell\,\uv$ yield $s\la\rv\ra_s=v_0\la\uv\ra_s-\zeta\la\rv\ra_s-(\xi-\zeta)\la\ell\uv\ra_s$ and $(s+4\dr+\xi)\la\ell\uv\ra_s=2\dr\la\rv\ra_s+v_0\la\uv\ra_s$, so that
\bea
\la\rv\ra_s=\frac{v_0\uv_0\,(s+4\dr+\zeta)}{(s+\dr)\big[(s+\zeta)(s+4\dr+\xi)+2\dr(\xi-\zeta)\big]}\,.
\label{SM:eq:rav_general}
\eea
The corresponding transient rates are
\bea
\gamma_\mp=\frac12\left[4\dr+\xi+\zeta\mp\sqrt{16\dr^2+(\xi-\zeta)^2}\right],
\eea
with $\gamma_-$ the slow rate.
For frozen orientation, $\la\rv\ra_s=v_0\uv_0/[s(s+\xi)]$ for every $\zeta$, so the quasi-steady mean position $a\,\uv_0$ is independent of $\muperp$.
The removable singularity of Sec.~\ref{SM:sec:aniso} generalizes to the condition that the bracket vanishes at $s=-\dr$, i.e. $\gamma_-=\dr$,
\bea
3\dr^2-\dr(\xi+\zeta)-\xi\zeta=0
\quad\Longleftrightarrow\quad
\chi_{\rm res}=\frac{1+\lambda+\sqrt{(1+\lambda)^2+12\lambda}}{6}\,,
\label{SM:eq:chires}
\eea
which gives $\chi_{\rm res}=1/3$ for $\lambda=0$ and $\chi_{\rm res}=1$ for $\lambda=1$.

\medskip

\noindent
{\bf Mean-squared displacement:~}
From Eq.~\eqref{SM:eq:generator_bf}, $\mathcal{L}\ell^2=2v_0\ell-2\Gam\ell^2+2\dr c^2$ and $\mathcal{L}c^2=-2(\dr+\zeta)c^2+2\dr\ell^2$.
In Laplace space,
\bea
(s+2\Gam)\la\ell^2\ra_s=2v_0\la\ell\ra_s+2\dr\la c^2\ra_s\,,\qquad
(s+2\dr+2\zeta)\la c^2\ra_s=2\dr\la\ell^2\ra_s\,,
\eea
and $\la\rv^2\ra_s=\la\ell^2\ra_s+\la c^2\ra_s$ gives
\bea
\la\rv^2\ra_s=\frac{2v_0^2\,(s+4\dr+2\zeta)}{s\,(s+\Gam)\big[(s+2\dr+2\xi)(s+2\dr+2\zeta)-4\dr^2\big]}\,.
\label{SM:eq:r2_Laplace_general}
\eea
The transient rates are $\Gam$ and $2\dr+\xi+\zeta\pm\sqrt{4\dr^2+(\xi-\zeta)^2}$.
The square-root structure is controlled by the mobility contrast $\xi-\zeta$ and disappears continuously as $\muperp\to\mupara$.
For $\dr=0$, Eq.~\eqref{SM:eq:r2_Laplace_general} gives $\la\rv^2\ra=a^2(1-e^{-\xi t})^2$ for every $\zeta$: the quasi-steady plateau is unaffected by $\muperp$.
The steady-state values are
\bea
\la\rv^2\ra_{\rm st}=\frac{v_0^2\,(2\dr+\zeta)}{\Gam\,\Pi}\,,\qquad
\la(\uv\cdot\rv)^2\ra_{\rm st}=\frac{v_0^2\,(\dr+\zeta)}{\Gam\,\Pi}\,,\qquad
\Pi\equiv\dr(\xi+\zeta)+\xi\zeta\,.
\label{SM:eq:r2_general_st}
\eea

\medskip

\noindent
{\bf Steady-state fourth moment:~}
Solving the degree-3 and degree-4 systems of Eq.~\eqref{SM:eq:master} in steady state gives
\bea
\la\rv^4\ra_{\rm st}=\frac{v_0^4\,N_4}{\Gam\,\Pi\,\big[3\dr^2+8\dr\xi+\xi^2+2\zeta(\dr+\xi)\big]\,M_4}\,,
\label{SM:eq:r4_general_st}
\eea
with
\bea
N_4 &=& 96\dr^4+\dr^3(40\xi+184\zeta)+\dr^2(6\xi^2+56\xi\zeta+116\zeta^2)
+\dr\,\zeta\,(4\xi^2+25\xi\zeta+27\zeta^2)+\zeta^2(\xi^2+3\xi\zeta+2\zeta^2)\,,\nn\\
M_4 &=& 4\dr^2(\xi+\zeta)+\dr(\xi^2+8\xi\zeta+\zeta^2)+\xi\zeta(\xi+\zeta)\,,
\eea
and $\mathcal{K}^{\rm st}=\la\rv^4\ra_{\rm st}/(2\la\rv^2\ra_{\rm st}^2)-1$.
For $\zeta=0$ these reduce to Eqs.~\eqref{SM:eq:r4_aniso_st} and~\eqref{SM:eq:alpha2_aniso}, and for $\zeta=\xi$ to Eqs.~\eqref{SM:eq:r4_iso_st} and~\eqref{SM:eq:alpha2_iso}.
Note that $M_4$ is an algebraic combination of rates and should not be confused with the characteristic polynomial $Q_4(s)$ of Eq.~\eqref{SM:eq:Qn}.
Figure~\ref{SM:fig:S2}(b) compares $\mathcal{K}^{\rm st}$ with simulations for several $\lambda$.

\medskip

\noindent
{\bf High-persistence limit:~}
Consider $\chi\to0$ at fixed
\bea
\epsilon\equiv\frac{\zeta}{\dr}=\frac{\lambda}{\chi}\,,
\eea
which compares transverse mechanical relaxation with rotational relaxation.
In Eq.~\eqref{SM:eq:bodyframe}, $\ell$ relaxes at rate $\Gam\simeq\xi$ to $a$, and its noise, of variance $\sim\dr\la c^2\ra/\xi$, is negligible.
Setting $\ell=a$ in the equation for $c$ gives an Ornstein--Uhlenbeck process,
\bea
dc=-(\dr+\zeta)\,c\,dt-\sqrt{2\dr}\,a\,dW\,,
\eea
whose restoring rate contains $\dr$ itself (from the It\^o drift of $\uv_\perp$).
Its stationary distribution is Gaussian with variance $a^2\dr/(\dr+\zeta)=a^2/(1+\epsilon)$.
With $r^2=a^2+c^2$ this yields
\bea
\la\rv^2\ra_{\rm st}\to a^2\,\frac{2+\epsilon}{1+\epsilon}\,,\qquad
\mathcal{K}^{\rm st}\to-\frac{\epsilon^2+4\epsilon+2}{2(2+\epsilon)^2}\,,
\label{SM:eq:K_hp}
\eea
in agreement with the $\chi\to0$ limit of Eqs.~\eqref{SM:eq:r2_general_st} and~\eqref{SM:eq:r4_general_st} at fixed $\epsilon$.
The kurtosis interpolates monotonically between the anisotropic value $-1/4$ ($\epsilon=0$) and the isotropic value $-1/2$ ($\epsilon\to\infty$).
Changing variables from $\tilde c=c/a$ to $\tilde r=r/a=\sqrt{1+\tilde c^2}$, with two values of $\tilde c$ per $\tilde r$, gives the radial distribution
\bea
P(\tilde r)=\sqrt{\frac{2(1+\epsilon)}{\pi}}\;
\frac{\tilde r}{\sqrt{\tilde r^2-1}}\;
\exp\!\left[-\frac{(1+\epsilon)(\tilde r^2-1)}{2}\right]\Theta(\tilde r-1)\,.
\label{SM:eq:P_hp}
\eea
The region $\tilde r<1$ is inaccessible, the distribution has an integrable divergence at $\tilde r\to1^+$, and its weight lies entirely outside the ring.
At finite $\chi$ the divergence is smoothed by the fluctuations of $\ell$.
Equation~\eqref{SM:eq:P_hp} agrees with simulations without adjustable parameters [Fig.~\ref{SM:fig:S2}(c)].

\medskip

\noindent
{\bf Slow-sector scaling forms:~}
The same separation gives closed forms for the full time dependence on the slow timescale.
For $\dr,\zeta\ll\xi$ and $s=\mathcal{O}(\dr)$, the bracket in Eq.~\eqref{SM:eq:rav_general} reduces to $\xi(s+2\dr+\zeta)$ and that in Eq.~\eqref{SM:eq:r2_Laplace_general} to $2\xi(s+2\dr+2\zeta)$, so that
\bea
\la\rv\ra_s\to\frac{a\,(s+4\dr+\zeta)}{(s+\dr)(s+2\dr+\zeta)}\,,\qquad
\la\rv^2\ra_s\to\frac{a^2\,(s+4\dr+2\zeta)}{s\,(s+2\dr+2\zeta)}\,,
\eea
whose inverse transforms are
\bea
\la\rv\ra\cdot\uv_0&=&\frac{a}{1+\epsilon}\Big[(3+\epsilon)\,e^{-\dr t}-2\,e^{-(2\dr+\zeta)t}\Big],
\label{SM:eq:slow_mean}\\
\la\rv^{2}\ra&=&\frac{a^2}{1+\epsilon}\Big[(2+\epsilon)-e^{-2(\dr+\zeta)t}\Big].
\label{SM:eq:slow_msd}
\eea
The long-time limits reproduce Eq.~\eqref{SM:eq:K_hp}.
For $\epsilon=0$, Eq.~\eqref{SM:eq:slow_mean} becomes $a(3e^{-\dr t}-2e^{-2\dr t})$, which exceeds the quasi-steady plateau and peaks at
\bea
\la\rv\ra\cdot\uv_0\big|_{\rm max}=\frac{9a}{8}\qquad\text{at}\qquad \dr t=\ln(4/3)\,,
\eea
whereas the isotropic mean in the same limit is $a\,e^{-\dr t}$ and decays monotonically.
The $12.5\%$ overshoot visible in Fig.~\ref{fig2}(a) of the main text is therefore exact.
Equation~\eqref{SM:eq:slow_msd} shows that the second relaxation step has height $a^2/(1+\epsilon)$ and occurs at the single rate $2(\dr+\zeta)$, which accounts for the curves in Fig.~\ref{SM:fig:S2}(a).
Both forms carry relative corrections of order $\chi$.

\medskip

\noindent
{\bf Physical picture of the plateau:~}
For $\dr=0$ and $\zeta=0$, Eq.~\eqref{SM:eq:bodyframe} has a line of marginally stable fixed points, $\ell=a$ with arbitrary $c$.
Starting from the origin, the particle reaches $(\ell,c)=(a,0)$ within $\tau_{\rm mech}=1/\xi$ and remains there, which produces the quasi-steady plateaus with vanishing fluctuations.
Rotational diffusion then displaces the particle along the neutral line on the scale $\tau_{\rm rot}=1/\dr$, and the MSD rises from $a^2$ to $2a^2$.
A finite $\zeta$ makes the line attracting, reducing the second step to $a^2/(1+\epsilon)$ [Fig.~\ref{SM:fig:S2}(a)].
In the isotropic case the ring is attracting in all directions, rotation moves the particle rigidly along it, and the MSD shows a single step.

\medskip

\noindent
{\bf Accuracy of the perfectly anisotropic kurtosis:~}
The criterion $\muperp k\lesssim\dr$ follows from Eq.~\eqref{SM:eq:K_hp} alone and guarantees $\mathcal{K}^{\rm st}\ge-7/18\simeq-0.389$, which already separates the anisotropic steady state from the isotropic value $-1/2$.
Recovering the value $-1/4$ itself is more demanding, because the leading finite-$\chi$ correction is not small in the region where $\epsilon$ is.
Expanding Eqs.~\eqref{SM:eq:r2_general_st} and~\eqref{SM:eq:r4_general_st} at fixed $\epsilon$,
\bea
\mathcal{K}^{\rm st}=-\frac{\epsilon^2+4\epsilon+2}{2(2+\epsilon)^2}
-\frac{6\epsilon^2+22\epsilon+13}{(1+\epsilon)(2+\epsilon)^2}\,\chi+\mathcal{O}(\chi^2)\,,
\label{SM:eq:Kcorr}
\eea
so that $\mathcal{K}^{\rm st}\simeq-\tfrac14-\epsilon/4-13\chi/4$ for $\epsilon,\chi\ll1$.
Since $\epsilon=\lambda/\chi$, the two corrections cannot be made small simultaneously.
Maximizing over $\chi$ at fixed $\lambda$ gives the height and location of the intermediate shoulder,
\bea
\chi_{\rm sh}=\sqrt{\lambda/13}\,,\qquad
\mathcal{K}^{\rm st}_{\rm sh}=-\frac14-\frac{\sqrt{13}}{2}\sqrt{\lambda}\,,
\label{SM:eq:shoulder}
\eea
in close agreement with the exact local maxima in Table~\ref{SM:tab:chistar} for $\lambda\lesssim10^{-3}$.
The plateau value $-1/4$ is therefore approached only as $\sqrt{\muperp/\mupara}$, and is resolved within a tolerance $\delta$ only for
\bea
\frac{\muperp}{\mupara}\lesssim\frac{4\delta^2}{13}\,,
\eea
that is, $\lambda\lesssim1.2\times10^{-4}$ for $\delta=0.02$.
At $\lambda=10^{-3}$ the exact local maximum is $-0.302$ (asymptotic shoulder $-0.307$); at $\lambda=10^{-2}$ the exact local maximum is $-0.382$, whereas the asymptotic formula gives $-0.430$ and is already inaccurate.

\medskip

\noindent
{\bf Robustness of the non-monotonic kurtosis:~}
Table~\ref{SM:tab:chistar} lists the extrema of $\mathcal{K}^{\rm st}(\chi)$ obtained from Eqs.~\eqref{SM:eq:r2_general_st} and~\eqref{SM:eq:r4_general_st}.
For $\lambda>0$ the strict limit $\chi\to0$ always gives $-1/2$, and the value $-1/4$ becomes an intermediate asymptote on the window $\lambda\ll\chi\ll1$, with the quantitative accuracy set by Eq.~\eqref{SM:eq:shoulder}.
A local maximum appears at $\chi\simeq\chi_{\rm sh}$, followed by the minimum near $\chi^*$.
As $\lambda$ increases, the minimum moves to smaller $\chi$ and deepens, and the two extrema merge at
\bea
\lambda_c\simeq0.2662\,,
\eea
beyond which $\mathcal{K}^{\rm st}$ is monotonic in $\chi$, as in the isotropic case.
The merging occurs at $\chi\simeq0.0879$, $\mathcal{K}^{\rm st}\simeq-0.4989$.
The singular point $\chi_{\rm res}$ of Eq.~\eqref{SM:eq:chires} moves in the opposite direction, which confirms that it is unrelated to $\chi^*$.
In summary, the two-step MSD and the enhanced steady-state MSD hold whenever $\muperp k\lesssim\dr$, the kurtosis minimum survives for $\muperp/\mupara<\lambda_c$, and the saturation value $-1/4$ requires the much stronger condition $\muperp/\mupara\lesssim10^{-4}$.
For slender bodies in a bulk Newtonian fluid, $\lambda\simeq1/2>\lambda_c$~\cite{Doi1986}, so the non-monotonic kurtosis and the high-persistence weight outside the ring require stronger anisotropy than hydrodynamics alone provides, as realized by wheeled, gliding, or structured-media particles.

\begin{table}[h]
\caption{Extrema of $\mathcal{K}^{\rm st}(\chi)$ for finite transverse mobility, $\lambda=\muperp/\mupara$. The local maximum exists only for $0<\lambda<\lambda_c\simeq0.266$. The last two columns give the removable singularity of $\la\rv\ra$, Eq.~\eqref{SM:eq:chires}, and the asymptotic shoulder height of Eq.~\eqref{SM:eq:shoulder}.}
\begin{ruledtabular}
\begin{tabular}{ccccccc}
$\lambda$ & $\chi^*$ (min) & $\mathcal{K}^{\rm st}(\chi^*)$ & $\chi$ (local max) & $\mathcal{K}^{\rm st}$ (local max) & $\chi_{\rm res}$ & $-\tfrac14-\tfrac{\sqrt{13}}{2}\sqrt{\lambda}$ \\
\hline
0        & 0.3318 & $-0.4643$ & --      & --        & 0.3333 & $-0.2500$ \\
$10^{-5}$& 0.3317 & $-0.4643$ & 0.00088 & $-0.2556$ & 0.3333 & $-0.2557$ \\
$10^{-4}$& 0.3316 & $-0.4643$ & 0.00277 & $-0.2675$ & 0.3335 & $-0.2680$ \\
$10^{-3}$& 0.3302 & $-0.4646$ & 0.00871 & $-0.3016$ & 0.3347 & $-0.3070$ \\
0.01     & 0.3167 & $-0.4676$ & 0.0269  & $-0.3819$ & 0.3463 & $-0.4303$ \\
0.05     & 0.2651 & $-0.4785$ & 0.0558  & $-0.4575$ & 0.3925 & -- \\
0.1      & 0.2148 & $-0.4876$ & 0.0720  & $-0.4824$ & 0.4421 & -- \\
0.2      & 0.1384 & $-0.4966$ & 0.0822  & $-0.4964$ & 0.5266 & -- \\
\end{tabular}
\end{ruledtabular}
\label{SM:tab:chistar}
\end{table}

\end{document}